# ARTICLE

# Quantum mechanics as classical statistical mechanics with an ontic extension and an epistemic restriction

Agung Budiyono[1,2] & Daniel Rohrlich[2]

Where does quantum mechanics part ways with classical mechanics? How does quantum randomness differ fundamentally from classical randomness? We cannot fully explain how the theories differ until we can derive them within a single axiomatic framework, allowing an unambiguous account of how one theory is the limit of the other. Here we derive non-relativistic quantum mechanics and classical statistical mechanics within a common framework. The common axioms include conservation of average energy and conservation of probability current. But two axioms distinguish quantum mechanics from classical statistical mechanics: an "ontic extension" defines a nonseparable (global) random variable that generates physical correlations, and an "epistemic restriction" constrains allowed phase space distributions. The ontic extension and epistemic restriction, with strength on the order of Planck's constant, imply quantum entanglement and uncertainty relations. This framework suggests that the wave function is epistemic, yet it does not provide an ontic dynamics for individual systems.

[1] Edelstein Center, Hebrew University of Jerusalem, Jerusalem 91904, Israel. [2] Department of Physics, Ben-Gurion University of the Negev, Beersheba 8410501, Israel. Correspondence and requests for materials should be addressed to A.B. (email: agungbymlati@gmail.com) or to D.R. (email: rohrlich@bgu.ac.il)

**D**o pure quantum states correspond one-to-one to real physical states? The claim that they do not—that they represent states of incomplete knowledge or information about the actual physical states—dates back to the first years of quantum mechanics[1,2]. In recent decades, this claim has inspired a research program[3–12] taking distinct pure quantum states to be compatible with a single physical state. That is, even pure quantum states in quantum mechanics could play a role similar to phase space distributions in classical statistical mechanics. Following refs. [7,13,14], we shall call the underlying physical state the "ontic state" and the probability distribution over the ontic states associated with a given quantum state the "epistemic state".

This program prompts the following question: Can we reconstruct quantum mechanics as a clear, physical modification of classical statistical mechanics? (An answer could be of practical interest toward understanding the physical resource responsible for the advantages of quantum over classical information protocols.) To answer this question, we note two essential features of quantum mechanics that distinguish it from classical mechanics: entanglement and the uncertainty principle. Can these two features lead us toward a reconstruction of quantum mechanics from classical statistical mechanics? Our only hope for a positive answer is to first adapt entanglement and the uncertainty principle to the framework of classical statistical mechanics. There have been attempts to explain (parts of) quantum mechanics starting from classical statistical models by assuming a fundamental restrictions on the class of possible epistemic states that can be prepared[6,9–12,15]. Such an "epistemic restriction" captures, to an extent, the uncertainty principle. Indeed, it is remarkable that this approach has been shown to reproduce a substantial portion of quantum phenomena traditionally judged incompatible with any classical world view.

In general, a model or theory that assumes that physical reality exists independent of measurement is called an "ontological" (or "ontic") model[13,14,16]. For example, Bohmian mechanics[17] is an "ontic extension" of classical mechanics: it posits a physically real or ontic and in general nonseparable wave function defined in a multi-dimensional configuration space, satisfying the Schrödinger equation, that guides the dynamics of the particles. (For an alternative nomological interpretation of the wave function within Bohmian mechanics, see ref. [18].) Conversely, Pusey, Barrett, and Rudolph have shown that any ontological model of quantum mechanics in which the wave function is not physically real must violate a statistical independence requirement called preparation independence[19]. Of course, such a model must also violate Bell's inequalities[20] and satisfy the Bell–Kochen–Specker contextuality theorem[15,21] and its generalization[16].

Here we derive nonrelativistic quantum mechanics (without quantum spin) in a way closely paralleling the derivation of classical statistical mechanics. That is, we derive the two theories within a common axiomatic framework, imposing conservation of average energy and probability current. Within this framework, two axioms distinguish quantum from classical statistical mechanics. One axiom creates an "ontic extension" in the form of "a global-nonseparable random variable;" the other axiom imposes a specific "epistemic restriction" on what probability distributions of momenta can be prepared, given a distribution of conjugate positions, and vice versa. (See Eqs. (5) and (6) below.) We obtain the mathematics and rules of quantum mechanics in a complex Hilbert space from this model as we average over the nonseparable random variable. We find that the ontic extension and epistemic restriction are together deeply related to two distinctive features of quantum mechanics: entanglement and the uncertainty principle, which arguably are responsible for the classically puzzling features of microscopic phenomena[22]. We conjecture and argue that, unlike Bohmian mechanics, the wave function in our ontological model is not physically real; what is real is a nonseparable, global random variable.

## Results

**Classical statistical mechanics of ensemble of trajectories.** Let us start from the conventional classical statistical mechanics of ensembles of trajectories. We work within the Hamilton–Jacobi formalism[23], proposing a set of specific modifications. One of the reasons we take the Hamilton–Jacobi formalism as our starting point is that in some formal classical limit, the Schrödinger equation (in the position representation) reduces to the Hamilton–Jacobi equation, with the quantum phase reducing to Hamilton's principal function. This limit is formally nontrivial; see for example the discussion in ref. [6]. We will argue that this formal limit can be extended conceptually, i.e., that Schrödinger's equation can likewise be interpreted as describing the dynamics of an ensemble of trajectories. Indeed, we will derive both fundamental equations within one axiomatic framework.

Let us consider a general many-particle system with $N$ degrees of freedom, in a spatial configuration denoted as $q = (q_1, \dots, q_N) \in \mathbb{R}^N$. Let $t$ denote the time parameterizing its evolution. Within the Hamilton–Jacobi formalism, the central result of classical statistical mechanics, based on the principle of least action, is that there is a differentiable real-valued scalar function of the configuration and time $S_C(q;t) : \mathbb{R}^N \times \mathbb{R} \mapsto \mathbb{R}$ with the dimensions of action—Hamilton's principal function—so that the momentum field $p = (p_1, \dots, p_N)$ canonically conjugate to $q$ is given by its gradient as:

$$p_i(q;t) = \partial_{q_i} S_C, \tag{1}$$

$i = 1, \dots, N$; and the time evolution of $S_C(q;t)$ is generated by the classical Hamiltonian $H(p,q)$ obeying the Hamilton–Jacobi equation:

$$-\partial_t S_C = H(p, q). \tag{2}$$

Equations (1) and (2) describe the dynamics of an ensemble of trajectories characterized by a single function $S_C$. A choice of $q$ at a single $t$ specifies the dynamics of a single trajectory.

As long as we consider a classical Hamiltonian up to second order in momentum, there is another way to derive the Hamilton–Jacobi Eq. (2), without resort to the least action principle. The derivation, given in Theorem 1 below, will provide physical insight for our reconstruction of quantum mechanics. Suppose we are given a continuous and deterministic momentum field in configuration space $p_i = p_i(q;t)$, $i = 1, \dots, N$. Let us construct a differentiable function $S_C(q;t)$ such that its spatial gradient at $(q;t)$ is precisely equal to $p$ as prescribed by Eq. (1). Next, let $\rho(q;t)$ denote the probability distribution over the configurations $q$ at time $t$. The phase space distribution, given a pair of functions $S_C$ and $\rho$, can then be written as $P(p, q|S_C, \rho) = P(p|q, S_C)\rho(q)$, where the conditional probability distribution of $p$ given $q$ and $S_C$ reads, noting Eq. (1),

$$P(p|q, S_C) = \prod_{i=1}^{N} \delta(p_i - \partial_{q_i} S_C); \tag{3}$$

time is implicit. For a pair $S_C$ and $\rho$, the ensemble (phase space) average of any physical quantity $\mathcal{O}(p,q) : \Omega \mapsto \mathbb{R}$, where $\Omega \equiv \mathbb{R}^{2N}$ is the phase space, is thus expressible as $\langle \mathcal{O} \rangle_{\{S_C, \rho\}} \equiv \int \mathrm{d}q \mathrm{d}p \, \mathcal{O}(p, q) P(p, q|S_C, \rho) = \int \mathrm{d}q \, \mathcal{O}(\partial_q S_C, q) \rho(q)$ with $\mathrm{d}q \equiv \mathrm{d}q_1 \dots \mathrm{d}q_N$, $\mathrm{d}p \equiv \mathrm{d}p_1 \dots \mathrm{d}p_N$.

How does the phase space distribution evolve over time? It is clear from the above construction that it is determined by the time evolution of $S_C$ and $\rho$, since this pair yields the phase space distribution. The time evolution of $\rho$ depends, via the continuity equation, on the velocity field, which in turn is determined by $S_C$ via Eq. (1). Hence, to obtain the time evolution of the phase space distribution, it suffices to know the dynamical equation for $S_C(q;t)$. Of course, we need constraints, characterizing the dynamics of the ensemble trajectories in phase space, to single out the dynamics of $S_C(q;t)$. The following theorem, proved in "Methods" subsection "Proof of Theorem 1", then applies:

**Theorem 1.** Consider an ensemble of trajectories satisfying Eq. (1) or equivalently Eq. (3). For a classical Hamiltonian $H(p,q)$ up to second order in momentum, the constraint that the ensemble of trajectories conserves the probability current and average energy implies a unique dynamics for $S_C(q,t)$, given by the Hamilton–Jacobi equation, Eq. (2):

$$-\partial_t S_C = H(p, q).$$

The Hamilton–Jacobi equation immediately implies the usual Liouville equation describing the time evolution of the phase–space distribution.

In the course of the paper, we shall not explicitly use Theorem 1. However, as mentioned earlier, it provides physical insight into, and thus anticipates, Theorem 3 below, which states that the Schrödinger equation can likewise be derived by imposing conservation of average energy and probability current. The only difference is that rather than constraining the ensemble of trajectories to satisfy Eq. (3), which is necessary for the derivation of the Hamilton–Jacobi equation, for deriving the Schrödinger equation we need a different class of trajectories, satisfying certain constraints to be discussed in the next subsection (cf. Eq. 6). On the one hand, this line of reasoning will provide a smooth formal and conceptual quantum-classical correspondence. On the other hand, we expect it to also tell us what really distinguishes quantum mechanics from classical statistical mechanics.

Now we make two remarks on classical statistical mechanics. First, in classical mechanics, the ontic (physical/microscopic) state is completely specified by the values of a pair of canonically conjugate variables $(p,q) \in \Omega$, or a point in the phase space $\Omega$. In a system having many subsystems, the space of the ontic states of the whole system is thus always "separable" into the Cartesian product of spaces of subsystems: $\Omega = \Omega_1 \times \ldots \times \Omega_N$. Also, the time evolution of the ontic state is deterministic. Second, the epistemic (macroscopic) state in classical statistical mechanics is given by the probability distribution over the phase space. One can then explicitly see from Eq. (1) or (3) that in classical statistical mechanics there is an "epistemic" (statistical) freedom to choose an arbitrary momentum field $p(q)$ consistent with a given $\rho(q)$. In other words, we are free to select the conditional probability distribution over the momentum independently of the distribution over the canonically conjugate position:

$$P(p|q, S_C, \rho) = P(p|q, S_C). \tag{4}$$

Conversely, given a momentum field $p(q)$, there is a freedom to prepare an ensemble of trajectories compatible with $p(q)$ with arbitrary $\rho(q)$. That is, each trajectory in the momentum field can be weighted arbitrarily. We show below how to derive quantum mechanics by giving up these two basic features of classical statistical mechanics.

**Microscopic ontic extension and epistemic restriction.** Evidently, we need new physical axioms if we are to recover quantum mechanics. Here we introduce the following two innovations. First, we make an ontic extension by introducing a hypothetical global-nonseparable ontic variable $\xi$. It is real valued with dimensions of action, and depends on time—it is spatially uniform. By "global-nonseparable," we mean that at any time, two arbitrarily separated physical objects are subject to the same simultaneous value of $\xi$. We assume that $\xi$ fluctuates "randomly" on a microscopic time scale with a probability distribution at any time given by $\mu(\xi)$, so that each single run of an experiment, in an ensemble of identical experiments, is parameterized by an independent realization of $\xi$.

Second, we assume the following epistemic restriction on the possible phase space probability distributions that nature allows us to prepare. We assume that, given a momentum field, it is not possible to prepare an ensemble of trajectories compatible with the momentum field with arbitrary distribution over configurations, again denoted by $\rho(q;t)$. In other words, each trajectory in the momentum field can no longer be assigned arbitrary weight. In turn, the conditional distribution of $p$, given $q$ at time $t$, therefore depends on the choice of $\rho(q;t)$, so that we have:

$$P(p|q, \ldots, \rho) \neq P(p|q, \ldots) \tag{5}$$

(in contrast to Eq. 4). Hence, we shall consider a model which lacks the epistemic freedom of classical statistical mechanics.

Let us then assume that an ensemble of identical preparations (defined by the same set of macroscopic parameters) will generate a conditional probability distribution for the momenta which depends on $\rho$ in accord with Eq. (5) as follows:

$$P(p|q, \xi, S_Q, \rho) = \prod_{i=1}^{N} \delta\left(p_i - \left(\partial_{q_i} S_Q + \frac{\xi}{2} \frac{\partial_{q_i} \rho}{\rho}\right)\right); \tag{6}$$

time is implicit. Here $S_Q(q;t) : \mathbb{R}^N \times \mathbb{R} \mapsto \mathbb{R}$ is a real-valued scalar function with dimensions of action; it replaces $S_C(q;t)$ of the corresponding classical case (cf. Eq. 3). In general, $S_Q$ may have a different form from $S_C$: the time evolution of $S_Q$, as will be shown later, will have to satisfy a modified Hamilton–Jacobi equation with a $\rho$-dependent term. (See "Methods" subsections "Proof of Theorem 3" and "Schrödinger's equation for measurement of angular momentum"). The ensemble of trajectories must satisfy Eq. (6). Further on the consistency of the statistical interpretation of Eq. (6) is given in "Methods" subsection "Statistical interpretation of the epistemic restriction."

For a smooth macroscopic classical limit, Eq. (6) must reduce to the classical form of Eq. (3). Thus, the classical limit must have $|\partial_q S_Q| \gg |\xi/2| |\partial_q \rho / \rho|$, and $S_Q \to S_C$. This condition suggests that the fluctuations of $\xi$, characterizing the strength of both the ontic extension and the epistemic restriction, must be microscopic. Hence, let us take the universal mean and variance of $\xi$ to be

$$\overline{\xi} \equiv \int d\xi \, \xi \, \mu(\xi) = 0, \quad \sigma_{ns}^2 \equiv \overline{\xi^2} - \overline{\xi}^2 = \overline{\xi^2} = \hbar^2, \tag{7}$$

respectively. One then sees that the formal limit $|\xi| \to 0$ is equivalent to $\sigma_{ns} = \hbar \to 0$.

We have thus, in $\xi$, introduced a fundamental concept of "ontic nonseparability"[8,24,25], with strength given by the Planck constant. The nonseparability of $\xi$ will be shown in the next subsection to be necessary for determining the correlations among systems and the average interaction energy between two systems, and to obtain the correct (linear) Schrödinger equation governing interacting systems, generating quantum entanglement. The ontic extension introducing $\xi$ with statistics given in Eq. (7), and the epistemic restriction of Eq. (6), together distinguish the quantum from the classical world. Let us mention that a toy model combining an epistemic restriction with an ontic extension in the form of a "relational stochastic variable" was recently proposed in ref. [25] to address the Pusey–Barrett–Rudolph theorem[19]. Note that, since $\xi$ does not depend on $q$, also $\sigma_{ns} = \hbar$ does not depend on $q$. Below, we will further assume that $\sigma_{ns} = \hbar$ does not depend on time, either. We will see that the spatiotemporal neutrality of $\sigma_{ns} = \hbar$ is crucial for reconstructing quantum mechanics.

Let us illustrate the epistemic restriction of Eqs. (6) and (7) in one spatial dimension, as follows. First, a trajectory passing randomly through $q$ acquires a fluctuating momentum $p$ from $\xi$. The strength of the fluctuation of $p$ is proportional to the strength $\sigma_{ns} = \hbar$ of the fluctuation of $\xi$, as in Eq. (7), and to the normalized slope of $\rho(q)$. We exhibit the meaning of these fluctuations in two extreme cases that rule out any epistemic state that is sharp both in $p$ and $q$. First, suppose we fix $p$ to be $\bar{p}$, that is $P(p|q,\xi,S_Q,\rho) = \delta(p - \bar{p})$. Equation (6) implies that the term $\partial_q \rho / \rho$ on the right-hand side of Eq. (6) must vanish; otherwise $p$ would fluctuate because $\xi$ does. Then for $\partial_q \rho / \rho$ to vanish, $\rho(q)$ must not depend on $q$. Hence, any attempt to fix $p$ inevitably implies complete uncertainty about $q$. Conversely, suppose we fix $q$ at $\bar{q}$, i.e., $\rho(q) = \delta(q - \bar{q})$. Then the term $\partial_q \rho / \rho$ in Eq. (6) diverges, implying a random fluctuation in $p$ with infinite strength. Hence, sharp knowledge about $q$ implies completely ignorance about $p$.

As a simple, concrete example, let us consider an ensemble of trajectories in a one-dimensional space with a distribution of position given by a Gaussian $\rho(q)$ of a vanishing mean and a variance $\sigma_q^2$, namely $\rho(q) = \exp(-q^2/2\sigma_q^2)$ (up to normalization), and take $S_Q$ independent of $q$, for simplicity. Substituting into Eq. (6) and noting Eq. (7), we obtain the variance $\sigma_p^2$ of $p$ as $\sigma_p^2 = \hbar^2/4\sigma_q^2$ and an uncertainty relation $\sigma_q \sigma_p = \hbar/2$, showing that it is impossible to prepare an epistemic state sharp both in $q$ and $p$ (via squeezing along both $p$ and $q$ axes). This uncertainty relation will be derived in general in the next subsection.

The example suggests that our epistemic restriction of Eqs. (6) and (7) is closely related to the knowledge-balance principle[7], the uncertainty principle[11], and the principle of classical complementarity[12], adopted by Spekkens et al. as a fundamental epistemic restriction for reconstructing a significant part of quantum mechanics from the statistical theory of some classical models; all these principles, as well, assert that complete knowledge of both $q$ and $p$ is impossible. Note, however, that none of these restrictions leads to a full reconstruction of quantum mechanics.

We will show that our epistemic restriction allows derivation of the uncertainty relation postulated as an epistemic restriction in ref.[7]. Thus, we adopt the epistemic restriction of Eqs. (6) and (7) as an axiom for reconstructing quantum mechanics. It clearly shows that maximal knowledge is always incomplete[2]. However, unlike the models in refs. [7,11,12], we also introduce an ontic extension—a fundamentally nonseparable random variable $\xi$—that induces an epistemic restriction of strength $\sigma_{ns} = \hbar$. Moreover, the nonseparability of $\xi$ will prove to be crucial for generating quantum entanglement via interaction.

**Emergent quantum kinematics and dynamics**. We assume that physical quantities $\mathcal{O}(p,q)$ in our model are real-valued functions of phase space, $\mathcal{O}(p,q) : \Omega \mapsto \mathbb{R}$, having the same form as those in classical mechanics. Hence, they depend on $\xi$ only implicitly via $p$ as in Eq. (6). We further confine ourselves to physical quantities that are at most second order in momentum. The two theorems presented below then assert that averaging over the fluctuations of the random variable $\xi$ leads to the mathematics and rules of quantum kinematics and dynamics, including the dynamics of measurement interaction.

**Theorem 2**. Assume that an ensemble of trajectories satisfies the epistemic restrictions of Eqs. (6) and (7), where $\sigma_{ns} = \hbar$ is constant in space. The phase space (ensemble) average $\langle\mathcal{O}\rangle_{\{S_Q,\rho\}}$ of any physical quantity $\mathcal{O}(p,q)$ up to second order in $p$ is then

equal to the quantum mechanical expectation value $\langle\psi|\hat{\mathcal{O}}|\psi\rangle$:

$$\langle\mathcal{O}\rangle_{\{S_Q,\rho\}} \equiv \int dq d\xi dp\, \mathcal{O}(p,q) P(p|q,\xi,S_Q,\rho)\mu(\xi)\rho(q) = \left\langle\psi\left|\hat{\mathcal{O}}\right|\psi\right\rangle, \tag{8}$$

where $\hat{\mathcal{O}}$ is the Hermitian operator obtained by applying the Dirac canonical quantization scheme to $\mathcal{O}(p,q)$ with a specific ordering of operators, and the wave function $\psi(q;t) = \langle q|\psi\rangle$ is defined as:

$$\psi(q;t) \equiv \sqrt{\rho(q;t)}\exp(iS_Q(q;t)/\hbar). \tag{9}$$

See the proof of Theorem 2 in the "Methods" section. (Here we have assumed that the joint probability distribution over $(q,\xi)$ is factorizable: $P(\xi,q) = \mu(\xi)\rho(q)$. In general, it may not be, and one has instead $P(\xi,q) = \mu(\xi|q)\rho(q)$, where $\mu(\xi|q)$ is the conditional probability of $\xi$ given $q$. All the results of calculations in this paper still apply unmodified in this general case if we replace the averages over $\mu(\xi)$ with those over $\mu(\xi|q)$, as long as the first two moments of $\mu(\xi|q)$ are given by Eq. (7) independent of $q$.)

Note first that, from Eq. (9), Born's statistical interpretation of the wave function is valid by construction: $P(q|\psi) = \rho(q) = |\psi(q)|^2$. Moreover, from Eqs. (6) and (9), each pure quantum state $\psi$ is associated with a phase–space distribution conditioned on $\xi$, namely $P(p,q|\xi,\psi) = P(p|q,\xi,S_Q,\rho)\rho(q) = \prod_{i=1}^N \delta(p_i - (\partial_{q_i}S_Q + \xi\partial_{q_i}\rho/2p_i))\rho(q)$. These results show the statistical aspect of the wave function. Although a consistent statistical interpretation of Eq. (6) can be made (as argued in "Methods" subsection "Statistical interpretation of the epistemic restriction"), this observation does not allow us to conclude that the wave function within the model defined in Eq. (9) is purely statistical with no physical (ontic) role. As discussed in that subsection, to have such a purely statistical $\psi$, we need to show that Eq. (6) can be derived from the ontic dynamics of individual systems with transparent causation, leading "effectively" to a statistical correlation between $p$ and $\rho$, with no causal relation. We conjecture that this is indeed the case.

As can be seen explicitly from the proof of Theorem 2 in the "Methods" section, the fundamental "nonseparability" of $\xi$ is necessary for obtaining Eq. (8). Suppose instead that $\xi$ is separable into $N$ random variables $\xi = (\xi_1, ..., \xi_N)$, $\xi_i$ is associated with the $i$-th degree of freedom replacing $\xi$ in Eq. (6). Assume that they have vanishing average $\overline{\xi}_i = 0$, $i = 1, ..., N$, and there is "independent" pair $\xi_i$, $\xi_j$ for some $i \neq j$ so that they are uncorrelated $\overline{\xi_i\xi_j} = \overline{\xi}_i\overline{\xi}_j = 0$. Then, the last term in Eq. (27) associated with the pair of indices $i$, $j$ vanishes.

For a concrete simple example showing the crucial role of the nonseparability of $\xi$, let us consider a pair of particles and compute the ensemble average of $\mathcal{O} = p_1 p_2$. Using Eq. (6) with $\xi_i$, $i = 1, 2$, for each degree of freedom replacing $\xi$, one directly gets $\langle p_1 p_2\rangle_{\{S_Q,\rho\}} = \int dq\left[(\partial_{q_1}S_Q)(\partial_{q_2}S_Q) + \overline{\xi_1\xi_2}(\partial_{q_1}\rho)(\partial_{q_2}\rho)/4\rho^2\right]\rho$, where we have made use of $\overline{\xi}_1 = 0 = \overline{\xi}_2$. In the nonseparable case, namely $\xi_1 = \xi_2 = \xi$, this gives us the quantum expectation value: $\langle p_1 p_2\rangle_{\{S_Q,\rho\}} = \int dq\left[(\partial_{q_1}S_Q)(\partial_{q_2}S_Q) + \hbar^2(\partial_{q_1}\rho)(\partial_{q_2}\rho)/4\rho^2\right]\rho = \int dq\left[(\partial_{q_1}S_Q)(\partial_{q_2}S_Q) - \hbar^2(\partial_{q_1}\partial_{q_2}R_Q)/R_Q\right]\rho = \langle\psi|\hat{p}_1\hat{p}_2|\psi\rangle$, where $R_Q \equiv \sqrt{\rho}$ and we have used Eq. (7) in the first equality; the identity of Eq. (24) (see the proof of Theorem 2 in the "Methods" section), and partial integration to get the second equality; and the definition of the wave function of Eq. (9), and $\langle q_i'|\hat{p}_i|q_i\rangle \equiv -i\hbar\partial_{q_i}\delta(q_i' - q_i)$, to arrive at the last equality. If we instead assume that $\xi_1$ and $\xi_2$ are independent (thus uncorrelated) random variables, namely $\overline{\xi_1\xi_2} = \overline{\xi}_1\overline{\xi}_2 = 0$, then we get $\langle p_1 p_2\rangle_{\{S_Q,\rho\}} = \int dq (\partial_{q_1}S_Q)(\partial_{q_2}S_Q)\rho(q)$, which is just the value obtained in conventional classical statistical mechanics (identifying $S_Q$ with $S_C$).

Notice in particular that, in the above example, if $\xi$ were separable into a pair of independent random variables, the quantum correction term $\int dq\, \hbar^2 \left[ (\partial_{q_1}\rho)(\partial_{q_2}\rho)/4\rho^2 \right]\rho = -\int dq\, \hbar^2 \left[ (\partial_{q_1}\partial_{q_2}R_Q)/R_Q \right]\rho$ would be missing. We will show later that this term generates a quantity, the quantum potential of Bohmian mechanics, that is crucial for the description of two particles[17]. (See "Methods" subsection "Schrödinger's equation for measurement of angular momentum" for a concrete example.) In contrast to our model, Bohmian mechanics postulates a quantum potential, taking $\rho = |\psi|^2$ and $\psi$ physically real. (One can equivalently postulate that $\psi$ follows the Schrödinger equation.) It is well known that the quantum potential plays a decisive role in any realist account of quantum mechanics, and is commonly regarded as responsible for many classically puzzling features of microscopic world. In our model, the quantum correction term arises effectively from the epistemic restriction of Eqs. (6) and (7) underlying the kinematics of the ensemble of trajectories. Moreover, for many-particle systems, as shown above, the nonseparability of $\xi$ is indispensable for the emergence of the quantum correction.

Let us note that the quantity $\langle p_1 p_2 \rangle_{\{S_Q, \rho\}} = \langle \psi | \hat{p}_1 \hat{p}_2 | \psi \rangle$ above can be regarded either as the momentum "correlation" between two arbitrarily separated particles, or as proportional to the "average interaction energy" between the two particles, which, e.g., arises in the von Neumann's prescription for measurement interaction. Hence, the fluctuation of $\xi$ not only fixes the strength of the epistemic restriction as discussed in the previous subsection; the irreducible nonseparability of $\xi$ plays a crucial role in describing the correlation between two particles (or subsystems) and their interaction. This role will become more prominent in the derivation of Schrödinger's equation for many interacting subsystems, in Theorem 3 below, in which we show that the nonseparability of $\xi$ is crucial for obtaining the correct Schrödinger equation describing interactions, and hence for obtaining quantum entanglement. Otherwise, if $\xi$ is separable, one will instead get a classical Hamilton–Jacobi equation. (See "Methods" subsection "Schrödinger's equation for measurement of angular momentum").

As an important corollary of Theorem 2, substituting $(p - \langle p \rangle_{\{S_Q, \rho\}})^2$ for $\mathcal{O}$ in Eq. (8) yields the standard deviation $\sigma_p$ of $p$ in the ensemble of trajectories, and shows that $\sigma_p$ equals the quantum mechanical standard deviation $\sigma_{\hat{p}}$ of $\hat{p}$. Namely, $\sigma_p^2 \equiv \langle (p - \langle p \rangle_{\{S_Q, \rho\}})^2 \rangle_{\{S_Q, \rho\}} = \langle \psi | (\hat{p} - \langle \psi | \hat{p} | \psi \rangle)^2 | \psi \rangle \equiv \sigma_{\hat{p}}^2$. Likewise, the standard deviation $\sigma_q$ of $q$ equals the quantum mechanical standard deviation $\sigma_{\hat{q}}$ of $\hat{q}$, i.e., $\sigma_q^2 \equiv \langle (q - \langle q \rangle_{\{S_Q, \rho\}})^2 \rangle_{\{S_Q, \rho\}} = \langle \psi | (\hat{q} - \langle \psi | \hat{q} | \psi \rangle)^2 | \psi \rangle \equiv \sigma_{\hat{q}}^2$. Hence, the standard deviations $\sigma_p$, $\sigma_q$ of the ensemble of trajectories satisfying Eqs. (6) and (7) always formally satisfy the Heisenberg–Kennard uncertainty relation[26–28]:

$$\sigma_q \sigma_p = \sigma_{\hat{q}} \sigma_{\hat{p}} \geq \hbar/2. \tag{10}$$

An alternative derivation of this uncertainty relation, which does not refer to Eq. (8), but directly applies the epistemic restriction given by the pair of Eqs. (6) and (7), appears in "Methods" subsection "An alternative derivation of the uncertainty relation."

The uncertainty relation of Eq. (10) describes a constraint on the epistemic states that can be prepared, rather than on simultaneous values of position and momentum. A similar uncertainty relation, together with the maximum entropy principle, is used in the ontological model of ref. [11] to derive a simplified quantum mechanics, called Gaussian quantum mechanics, from classical statistical mechanics. Unlike ref. [11], however, we do not impose the principle of maximum entropy. Thus, we recover the non-Gaussian regime.

The next question is how the ensemble of trajectories in our model evolves with time and how the evolution transforms the corresponding phase space distribution. Equations (6) and (9) tell us that the time evolution of the phase–space distribution is determined by that of $\psi$. Moreover, it is also clear that any type of time evolution for $\psi$ will preserve the uncertainty relation of Eq. (10). What, then, is the dynamical equation governing $\psi$? In the classical case, as mentioned in Theorem 1, the Hamilton–Jacobi equation—and thus the Liouville equation—are obtained by imposing the requirement that the ensemble of trajectories conserves the average energy and probability current. To have a conceptually smooth classical correspondence, we want the same requirement to single out the dynamical equation for $\psi$. And it does, as follows:

**Theorem 3**. Consider an ensemble of trajectories satisfying Eqs. (6) and (7), where $\sigma_{ns} = \hbar$ is constant in space and time. Given a classical Hamiltonian $H(p,q)$ that is at most quadratic in $p$, and an ensemble of trajectories conserving the average energy and probability current, there is a unique time evolution for $\psi$ given by the (linear and unitary) Schrödinger equation:

$$i\hbar \frac{d}{dt} |\psi\rangle = \hat{H} |\psi\rangle, \tag{11}$$

where $\hat{H}$ is a Hermitian operator again having the same form as that obtained by applying the Dirac canonical quantization procedures to $H(p,q)$ with a specific ordering of operators. See the proof of Theorem 3 in the "Methods" section.

As a first corollary to Theorem 3, in the macroscopic classical physical regime $|\partial_q S_Q| \gg |\xi/2| |\partial_q \rho/\rho|$, we regain the dynamical equation governing classical statistical mechanics, the Hamilton–Jacobi equation. For a proof, recall that, in the limit, the epistemic restriction of Eq. (6) reduces to the conditional probability distribution over $p$ in classical statistical mechanics, given by Eq. (3); so the average energy defined as in Eq. (8) must also reduce to the value obtained in classical statistical mechanics. Accordingly, by Theorem 1, the Schrödinger equation of Eq. (11) in the position representation must reduce in this limit to the classical Hamilton–Jacobi equation of (2) where $S_Q \rightarrow S_C$.

As a second corollary, one can show that interaction in the past in general implies a non-factorizable (entangled) wave function. Consider the limiting classical case where one has, via Eqs. (1) and (2), the equality $S_C(q;t) = \int^{(q;t)} L\, dt$, where $L(q,\dot{q}) = p \cdot \dot{q} - H$ is the classical Lagrangian. Next, consider two subsystems with a configuration $q = (q_1, q_2)$. Assume that they interacted, so that there was an interval of time in the past during which the total Lagrangian was not (additively) decomposable into that of the two subsystems: $L(q, \dot{q}) \neq L_1(q_1, \dot{q}_1) + L_2(q_2, \dot{q}_2)$. It follows that also Hamilton's principal function does not, in general, decompose: $S_C(q;t) = \int^{(q;t)} L\, dt \neq S_{C_1}(q_1;t) + S_{C_2}(q_2;t)$. Accordingly, for a smooth classical limit, also $S_Q(q;t)$ must not, in general, decompose: $S_Q(q;t) \neq S_{Q_1}(q_1;t) + S_{Q_2}(q_2;t)$. The wave function defined in Eq. (9) is therefore, in general, non-factorizable: $\psi(q;t) \neq \psi_1(q_1;t)\psi_2(q_2;t)$.

Note that for the two interacting subsystems above, to get quantum entanglement via Schrödinger equation, the nonseparability of $\xi$ is crucial. If instead we assume that $\xi$ is separable into two independent random variables $\xi_1$ and $\xi_2$ with vanishing average $\overline{\xi}_1 = 0 = \overline{\xi}_2$, so that $\overline{\xi_1 \xi_2} = \overline{\xi}_1 \overline{\xi}_2 = 0$, we will not get the correct quantum mechanical entangled wave function. Instead, as shown above, the average interaction energy is given by the classical statistical mechanics value, rather than the quantum expectation value. Hence, imposing the principle of conservation of average energy leads, as Theorem 1 shows, to the Hamilton–Jacobi equation with a classical interaction Hamiltonian instead of the Schrödinger equation for interacting subsystems. (For a concrete example, see the discussion at the end of "Methods" subsection "Schrödinger's equation for the measurement of angular momentum".) Indeed in the ontological model, the nonseparability of $\xi$ is crucial for obtaining nonfactorizable (entangled) wave functions.

We obtain yet another corollary of Theorem 3 if we couple a system to a measuring device via the von Neumann measurement-interaction Hamiltonian $H_I = g\mathcal{O}_S p_\Sigma$, where $\mathcal{O}_S(p_S, q_S)$, the physical quantity of the system to be measured, is linear in the momentum $p_S$, $p_\Sigma$ is the momentum of the apparatus pointer, and $g$ is the coupling strength. (Measuring a physical quantity $\mathcal{O}_S(p_S, q_S)$ that is second order in momentum requires a different measurement interaction, to make $H_I$ altogether only second order in momentum.) We get the Schrödinger equation of Eq. (11) with the quantum Hamiltonian $\hat{H}_I = g\hat{\mathcal{O}}_S \hat{p}_\Sigma$. The "Methods" subsection "Schrödinger's equation for measurement of angular momentum" provides an example of deriving the Schrödinger equation with a measurement interaction to measure angular momentum. From this result, given that the particles in our model always have definite positions and momenta as in Bohmian mechanics, it follows that each single measurement run will yield an outcome given by one of the eigenvalues of $\hat{\mathcal{O}}_S$ with statistics following Born's rule; and that the "effective" wave function of the system after the measurement is given by the eigenfunction of $\hat{\mathcal{O}}_S$ associated with the measurement outcome. We derive this rule explicitly in "Methods" subsection "Derivation of Born's rule", and discuss Wallstrom's critique[29] of this program for reconstructing quantum mechanics.

## Discussion

Among attempts to clarify the meaning and foundations of quantum mechanics, and to pinpoint its place among possible theories, there has been much interest recently in deriving quantum mechanics from physically transparent axioms[30–41]. In the present paper, partly inspired by the successes of the research program of epistemically restricted classical statistical models[6,7,9–12,15,25], which reproduce some quantum phenomena usually regarded as classically inexplicable, we have attempted to provide axioms for quantum mechanics that closely parallel the axioms of classical statistical mechanics, i.e., axioms within the same conceptual framework. Specifically, as Theorem 1 and Theorem 3 show, the dynamics of classical statistical mechanics and of our ontological model of quantum mechanics (which correspond, respectively, to the Hamilton–Jacobi equation and the Schrödinger equation) follows directly from axioms of conservation of average energy and probability current.

What transforms classical statistical mechanics into quantum mechanics, in our model, is the structure of the space of ontic and epistemic states and the dynamics of the ontic states. While in conventional classical statistical mechanics the ontic state follows deterministic dynamics and the space of the ontic states is separable, in our model the ontic extension arises from a nonseparable random variable $\xi$. Moreover, while classical mechanics allows preparing an ensemble of trajectories with an arbitrary distribution of positions, independently of a given momentum field and vice versa, in our model, quantum mechanics emerges when this independence is partially sacrificed in accordance with the epistemic restriction of Eqs. (6) and (7). The epistemic restriction of Eq. (6) can be generalized to any pair of canonically conjugate variables. We thus claim that two outstanding and paradoxical features of quantum mechanics, entanglement and uncertainty relations[15,16,19–22], are fundamentally related to this ontic extension and epistemic restriction.

We have presented the epistemic restriction of Eqs. (6) and (7) as a novel objective-realist approach to the Heisenberg uncertainty principle. The Heisenberg uncertainty principle connects the standard deviations of position and momentum measurements outcomes, whereas our approach connects probability distributions for momenta with probability distributions for positions, independent of measurement. Moreover, unlike the Heisenberg uncertainty principle from which, to the best of our knowledge, no one has derived Schrödinger's equation, here we have shown that the epistemic restriction and axioms of conservation of average energy and probability current do imply Schrödinger's equation. In this sense, the epistemic restrictions of Eqs. (6) and (7) are more powerful than the Heisenberg uncertainty principle.

Within the ontological model, conventional classical statistical mechanics emerges in the deterministic and separable physical regime when $|\partial_q S_Q| \gg |\xi/2| \, |\partial_q \rho/\rho|$, so that Eq. (6) reduces to Eq. (3). In this limit, evidently the ontic extension and the epistemic restriction vanish smoothly and jointly (they stand or fall together). These features of the model are appealing in the context of the long-standing problem of the quantum-classical correspondence: trajectories do not emerge as approximations to a macroscopic classical world; rather, they are well defined even in the microscopic world. Thus, we can also regard Theorem 2 and Theorem 3 as a novel quantization scheme[12] that applies only to systems for which the Hamiltonian is at most second order in momentum. Note that unlike Dirac canonical quantization procedure, as shown in the proof of Theorem 2 (see "Methods"), our scheme yields a unique Hermitian operator $\hat{\mathcal{O}}$ with a specific ordering of operators. Moreover, while Dirac quantization procedure is mathematically inspired, our scheme is physically and conceptually motivated. While we have focused on particles, this quantization scheme might find direct application in linear quantum optics, with $(p, q)$ as the field quadratures.

Many important questions are left for future work. Whence the specific epistemic restriction of Eqs. (6) and (7)? How does the model account for the quantum phenomena that seem least compatible with classical mechanics[22]? The answers must include an explanation of well-known no-go theorems such as Bell's theorem[20], the Bell–Kochen–Specker contextuality theorem[15,21] and its generalization[16], and the recent Pusey–Barrett–Rudolph theorem[19]. Our model suggests tracing these nonclassical phenomena and no-go theorems to the ontic extension and epistemic restriction imposed on an otherwise-conventional classical-statistical mechanics. To address quantum paradoxes, it might be necessary to obtain the epistemic restriction of Eq. (6) from a deeper causal model for individual systems. (See the discussion in "Methods" subsection "Statistical interpretation of the epistemic restriction".) Our model may stimulate novel ideas for simulating quantum information processing, shed new light on the physical nature of Planck's constant, and suggest a natural and consistent extension of quantum mechanics. Finally, it is necessary to extend our model to include spin and, ultimately, to confront relativistic invariance.

## Methods
**Proof of Theorem 1**. **Theorem 1:** Consider an ensemble of trajectories satisfying Eq. (1) or equivalently Eq. (2). For a classical Hamiltonian $H(p,q)$ up to second order in momentum, the constraint that the ensemble of trajectories conserves the probability current and average energy implies a unique dynamics for $S_C(q,t)$, given by the Hamilton–Jacobi equation, Eq. (2):

$$-\partial_t S_C = H(p, q).$$

*Proof.* We shall prove the theorem by considering a simple example of a single particle in three-dimensional space. The proof for the general case is completely analogous. Let us consider a single particle with mass $m$ subjected to a time-independent scalar potential $V(q)$ and a vector potential $A(q) = (A_1, A_2, A_3)$. The

Hamiltonian thus reads

$$H(p,q) = \sum_{i=1}^{3} \frac{[p_i - A_i(q)]^2}{2m} + V(q). \quad (12)$$

From Eq. (12), the velocity field is related to momentum field as: $\dot{q}_i(q;t) \equiv \mathrm{d}q_i/\mathrm{d}t = \partial H/\partial p_i = (p_i - A_i)/m$, so that noting Eq. (1), we get $\dot{q}_i = (\partial_{q_i} S_C - A_i)/m$. Assuming that the probability density current is conserved (i.e., no creation or annihilation of trajectories), which is a natural assumption for a closed system, the probability density $\rho(q;t)$ of $q$ at time $t$ satisfies a continuity equation:

$$0 = \partial_t \rho + \sum_i \partial_{q_i}(\rho \dot{q}_i) = \partial_t \rho + \sum_i \frac{1}{m} \partial_{q_i}[\rho(\partial_{q_i} S_C - A_i)]. \quad (13)$$

On the other hand, from Eqs. (3) and (12), the average energy of the ensemble of trajectories characterized by the same $S_C(q;t)$ and $\rho(q;t)$ is:

$$\begin{aligned} \langle H \rangle_{\{S_C,\rho\}} &\equiv \int \mathrm{d}q \mathrm{d}p \, H(p,q) \mathbb{P}(p|q,S_C) \rho(q;t) \\ &= \int \mathrm{d}q \left[ \sum_i \frac{(\partial_{q_i} S_C - A_i)^2}{2m} + V \right] \rho. \end{aligned} \quad (14)$$

Next, differentiating Eq. (14) with respect to time gives $(\mathrm{d}/\mathrm{d}t)\langle H \rangle_{\{S_C,\rho\}} = \int \mathrm{d}q \{ (\partial_t \rho)[\sum_i (\partial_{q_i} S_C - A_i)^2/2m + V] + [\sum_i \rho(\partial_{q_i} S_C - A_i) \partial_{q_i} \partial_t S_C/m] \}$. Integrating by parts the last term on the right-hand side and using the continuity equation of (13), we obtain:

$$\frac{\mathrm{d}}{\mathrm{d}t}\langle H \rangle_{\{S_C,\rho\}} = \int \mathrm{d}q \, \partial_t \rho \left[ \sum_i \frac{(\partial_{q_i} S_C - A_i)^2}{2m} + V + \partial_t S_C \right]. \quad (15)$$

The above relation clearly shows that the average energy is conserved for any time, i.e., $(\mathrm{d}/\mathrm{d}t)\langle H(t) \rangle_{\{S_C,\rho\}} = 0$ for any $\partial_t \rho$, if and only if the term inside the bracket is vanishing

$$\partial_t S_C + \sum_i \frac{(\partial_{q_i} S_C - A_i)^2}{2m} + V = 0. \quad (16)$$

This equation is just the Hamilton–Jacobi equation of (2), as we see by noting Eqs. (1) and (12).

We obtained the Hamilton–Jacobi equation of Eq. (2) by positing the kinematics of Eq. (1) or equivalently Eq. (3) and imposing the principles of conservation of average energy and probability current. The "Methods" subsections "Proof of Theorem 3" and "Schrödinger's equation for measurement of angular momentum" show that the same axiomatic framework yields, instead, the Schrödinger equation, if one posits the alternative kinematics of Eq. (6) or equivalently Eq. (17) below.

**Statistical interpretation of the epistemic restriction.** We discuss the possible conceptual issue which may arise in the epistemic restriction of Eq. (6). First, Eq. (6) can be equivalently written as:

$$p_i = \partial_{q_i} S_C + \frac{\xi}{2} \frac{\partial_{q_i} \rho}{\rho}, \quad (17)$$

$i = 1, \ldots, N$. A similar momentum fluctuation is also postulated in ref. [42], but no specific relation with $\rho$ is proposed, and no introduction of a global-nonseparable variable as in our model. It seems from the above equation that the momentum $p$ associated with $q$ has a given $\xi$ is in part determined by the probability density $\rho(q)$ for $q$. How can it be? Initially, this formal relation between $p$ and $\rho$ might give the impression that the dynamics of the particle is being guided causally by $\rho$. But such an interpretation grants causal power to mere epistemic possibilities (the probability density $\rho(q)$), which is unacceptable from the standpoint of statistical mechanics. We avoid this bizarre interpretation by denying $\rho$ an ontic status as in Bohmian mechanics[17] (in which $\rho$ determines the energy density via a term called quantum potential), or as in the many interacting worlds interpretation[43] (which assumes that all possible alternative realities co-exist).

Instead, as discussed in "Results" subsection "Microscopic ontic extension and epistemic restriction," we interpret the relation between $p$ and $\rho$ in Eqs. (6) or (17) as describing a statistical constraint or correlation, rather than as a causal relation, between $p(q;\xi)$ and $\rho(q)$. Namely, given a momentum field $p(q;\xi)$, among all possible classes of ensembles of trajectories with different weighting given by the different probability densities $\rho(q)$ that are compatible with $p(q;\xi)$, we choose a specific one with $\rho(q)$ that satisfies the constraint given by Eq. (17) for some $S_Q(q)$. From Eq. (17) (or equivalently Eq. (6) and Eq. (7), the form of $S_Q$ is determined by the average of $p$ over $\xi$ as $\bar{p}_i(q) = \int \mathrm{d}\xi \, p_i(q;\xi) \mu(\xi) = \partial_{q_i} S_Q$, $i = 1, \ldots, N$. Recall that in the classical case, any form of $\rho(q)$ is allowed (each trajectory belonging to the given momentum field can be weighted arbitrarily); here we have sacrificed part of this freedom. To stress this (non-bizarre) interpretation, we have formulated the epistemic restriction as a delta-functional conditional probability density of $p$ given $q$, $\xi$, $S_Q$ and $\rho$ in Eq. (6), as well as in the equivalent form of direct relation among $p$, $q$, $\xi$, $S_Q$ and $\rho$ of Eq. (17). The correlation between $p(q;\xi)$ and the gradient of $\rho(q)$ does not imply causation.

As an example, let us suppose that we are given a one-dimensional momentum field $p(q;\xi) = -\xi q/2\sigma_q^2$, where $\sigma_q$ is a constant. Within the model, since $\bar{p} = 0 = \partial_q S_Q$, this is the case when $S_Q$ is independent of $q$; it may still depend on time. In classical statistical mechanics, one is free to prepare any ensemble of trajectories compatible with a given momentum field with any arbitrary weighting given by $\rho(q)$, i.e., any form of $\rho(q)$ is allowed. In our model, however, such epistemic freedom is no longer granted. Instead, among ensembles of trajectories following the given momentum field $p(q;\xi) = -\xi q/(2\sigma_q^2)$, we select one with $\rho(q)$ that satisfies the epistemic restriction. Inserting this momentum field into the epistemic restriction of Eq. (17) and noting that $S_Q$ is independent of $q$, we have to choose $\rho(q)$ to solve the following differential equation: $\partial_q \rho/\rho = -q/\sigma_q^2$. This equation yields a Gaussian distribution of $q$: $\rho(q) \sim \exp\left(-q^2/2\sigma_q^2\right)$. As shown in "Results" subsection "Microscopic ontic extension and epistemic restriction," assuming Eq. (7) makes this ensemble of trajectories automatically satisfy the uncertainty relation $\sigma_q \sigma_p = \hbar/2$.

Hence, within our model, Eqs. (6) and (17) have physical meaning only for ensembles of identically prepared systems, and not for any individual systems. To have a complete realistic model, we need to provide a dynamics for the time evolution of individual systems (comparable to Newton's equation or the Langevin equation), with transparent causal structure; in particular such an ontic dynamics must not grant $\rho$ a causal role. We do not provide it here, but we "conjecture" that such an ontic dynamics consistent with Eq. (6) exists. This dynamics must therefore lead "effectively" to the emergence of the statistical correlations between $p$ and $\rho$ given in Eq. (6). This conjecture on the existence of ontic dynamics for individual systems with no causal role of $\rho$, implies that the wave function within our model is "epistemic". Remarkably, as we show in "Results" subsection "Emergent quantum kinematics and dynamics," the explicit ontic dynamics of individual systems is not needed for deriving the formal-mathematical concepts and operational rules of quantum mechanics.

Markopoulou and Smolin[44] and Smolin[45] suggest a similar notion by arguing that the dependence of energy density in Nelson's stochastic mechanics[46] (which corresponds to the quantum potential in Bohmian mechanics and is thus an ontic variable) on the spatial derivative of $\rho(q)$ (an epistemic parameter) arises effectively in a cosmological model where quantum mechanics is an approximation that applies only to a subsystem of the universe. See also ref. [47] in which the dependence arises effectively due to interactions between the particle and a zero-point radiation field, after averaging over the latter.

We end this subsection by presenting two more simple examples of how, for a given a momentum field, to define an ensemble of trajectories that satisfies the epistemic restriction of Eq. (6) or (17). As a first example, let us suppose that we are given a spatially uniform one-dimensional momentum field $p = p_0$, independent of $q$ and $\xi$. Clearly in this case, an ensemble of trajectories compatible with the momentum field must have a sharp distribution of momentum: they must all have momentum $p = p_0$. Again, recall that in classical mechanics, we are free to prepare an ensemble of trajectories compatible with a given momentum field with arbitrary $\rho(q)$. (Each trajectory belonging to the momentum field can be assigned arbitrary weight.) By contrast, our model offers no such freedom. The ensemble of trajectories must satisfy the statistical restriction of Eq. (17).

First, inserting $p = p_0$ into Eq. (17) and averaging over $\xi$, one gets $\bar{p} = \partial_q S_Q = p_0$, which can be integrated to yield $S_Q(q) = p_0 q + f(t)$, where $f(t)$ depends only on the time $t$. Inserting this back into Eq. (17), we see that $\rho(q)$ must therefore satisfy the differential equation $\partial_q \rho/\rho = 0$, which gives a spatially uniform $\rho(q)$, i.e., $\rho$ does not depend on $q$. Hence, only a spatially uniform $\rho(q)$ is allowed, consistent with our earlier result derived in "Results" subsection "Microscopic ontic extension and epistemic restriction," that an ensemble of trajectories with a sharp distribution of momentum must have a completely uncertain position. In this case, noting Eq. (9), the corresponding wave function is thus given by a plane wave $\psi \sim \exp(ip_0 q/\hbar)$ in accordance with the quantum mechanical notion that a plane wave function describes a spatially uniform ensemble of particles with a sharp momentum.

For another example, consider a particle in a one-dimensional box of unit length, $-1/2 \leq q \leq 1/2$. Let us assume that we are given a random momentum field of the form $p(q;\xi) = -\xi \pi \sin(\pi q)/\cos(\pi q)$. Again, in the classical case, given a momentum field, one is free to prepare any ensemble of trajectories compatible with the momentum field with arbitrary $\rho(q)$. By contrast, in our model, given the momentum field, only $\rho(q)$ satisfying the epistemic restriction of Eq. (17) is allowed. Notice first that at the two boundaries of the box, i.e., $q = \pm 1/2$, the momentum field is infinite. However, we shall soon see that the allowed probability density $\rho(q)$ for the particle to reach the wall of the box, which satisfies the epistemic restriction of Eq. (17), vanishes, i.e., $\rho(\pm 1/2) = 0$.

Namely, since the average of the momentum over $\xi$ is vanishing, $0 = \bar{p} = \partial_q S_Q$, then $S_Q$ must be independent of $q$ (but may still depend on time). Noting this and inserting the momentum field into the epistemic restriction of Eq. (17), we find that the probability density of $q$ must satisfy the following differential equation: $\partial_q \rho/\rho = -2\pi \sin(\pi q)/\cos(\pi q)$. Integrating this equation, we get $\rho(q) = 2 \cos^2(\pi q)$, which is just the probability density of $q$ corresponding to the quantum mechanical ground state of a particle in the box. One can check that the above momentum field, with the corresponding probability distribution of the position, automatically satisfies the uncertainty relation as shown in general in the main text in Eq. (10) and also in "Methods" subsection "An alternative derivation of the uncertainty relation." In fact, calculating the variance of $q$, one directly gets $\sigma_q^2 = \int_{-1/2}^{1/2} \mathrm{d}q \, q^2 \rho(q)$ $= 2 \int_{-1/2}^{1/2} dq\, q^2 \cos^2(\pi q) = (\pi^2 - 6)/12\pi^2$. On the other hand, calculating the variance of $p$ one obtains $\sigma_p^2 = \int_{-1/2}^{1/2} dq \int d\xi\, p(q;\xi)^2 \mu(\xi) \rho(q)$

$= 2\pi^2 \hbar^2 \int_{-1/2}^{1/2} dq \sin^2(\pi q) = \pi^2 \hbar^2$ where we have used Eq. (7). Hence, one has $\sigma_q \sigma_p = \sqrt{(\pi^2 - 6)/3}\, \hbar/2 \geq \hbar/2$.

**Proof of Theorem 2. Theorem 2:** Assume that an ensemble of trajectories satisfies the epistemic restrictions of Eqs. (6) and (7), where $\sigma_{ns} = \hbar$ is constant in space. The phase space (ensemble) average $\langle \mathcal{O} \rangle_{\{S_Q, \rho\}}$ of any physical quantity $\mathcal{O}(p, q)$ up to second order in $p$ is then equal to the quantum mechanical expectation value $\langle \psi | \hat{\mathcal{O}} | \psi \rangle$:

$$\langle \mathcal{O} \rangle_{\{S_Q, \rho\}} \equiv \int dq\, d\xi\, dp\, \mathcal{O}(p, q) P(p|q, \xi, S_Q, \rho) \mu(\xi) \rho(q) = \langle \psi | \hat{\mathcal{O}} | \psi \rangle, \quad (18)$$

where $\hat{\mathcal{O}}$ is the Hermitian operator obtained by applying the Dirac canonical quantization scheme to $\mathcal{O}(p, q)$ with a specific ordering of operators, and the wave function $\psi(q; t) = \langle q | \psi \rangle$ is defined as:

$$\psi(q; t) \equiv \sqrt{\rho(q; t)} \exp(iS_Q(q; t)/\hbar). \quad (19)$$

*Proof.* Let us first calculate, within the ontological model developed in the main text, the phase space (ensemble) average of a general classical physical quantity up to second order in momentum:

$$\mathcal{O}(p, q) = (g^{ij}(q)/2)(p_i - A_i(q))(p_j - A_j(q)) + V(q), \quad (20)$$

where $g^{ij}(q) = g^{ji}(q)$, $A_j(q)$ and $V(q)$ are real-valued functions and summation over repeated indices are assumed. One must evaluate the following integral:

$$\langle \mathcal{O} \rangle_{\{S_Q, \rho\}} \equiv \int dq\, d\xi\, dp\, \mathcal{O}(p, q) P(p|q, \xi, S_Q, \rho) \mu(\xi) \rho(q). \quad (21)$$

First, inserting Eqs. (6) and (20) into Eq. (21) one directly obtains, after a trivial integration over $p$,

$$\langle \mathcal{O} \rangle_{\{S_Q, \rho\}} = \int dq\, d\xi \left[ \frac{g^{ij}}{2} \left( \partial_{q_i} S_Q - A_i + \frac{\xi}{2} \frac{\partial_{q_i} \rho}{\rho} \right) \left( \partial_{q_j} S_Q - A_j + \frac{\xi}{2} \frac{\partial_{q_j} \rho}{\rho} \right) + V \right] \mu(\xi) \rho(q). \quad (22)$$

Expanding the multiplication in the bracket, integrating over $\xi$ and noting Eq. (7), one gets

$$\langle \mathcal{O} \rangle_{\{S_Q, \rho\}} = \int dq \left[ \frac{g^{ij}}{2} \left( \partial_{q_i} S_Q - A_i \right) \left( \partial_{q_j} S_Q - A_j \right) + V + \frac{\hbar^2}{4} \frac{g^{ij}}{2} \left( \frac{\partial_{q_i} \rho}{\rho} \right) \left( \frac{\partial_{q_j} \rho}{\rho} \right) \right] \rho. \quad (23)$$

Now let us proceed to evaluate the last term on the right-hand side of Eq. (23). Using the following mathematical identity

$$\frac{1}{4} \left( \frac{\partial_{q_i} \rho}{\rho} \right) \left( \frac{\partial_{q_j} \rho}{\rho} \right) = -\frac{\partial_{q_i} \partial_{q_j} R_Q}{R_Q} + \frac{1}{2} \frac{\partial_{q_i} \partial_{q_j} \rho}{\rho}, \quad (24)$$

where $R_Q \equiv \sqrt{\rho}$, we first have

$$I \equiv \int dq\, \frac{\hbar^2}{2} \frac{g^{ij}}{4} \left( \frac{\partial_{q_i} \rho}{\rho} \right) \left( \frac{\partial_{q_j} \rho}{\rho} \right) \rho = -\int dq\, \frac{\hbar^2}{2} \left( g^{ij} \frac{\partial_{q_i} \partial_{q_j} R_Q}{R_Q} - \frac{g^{ij}}{2} \partial_{q_i} \partial_{q_j} \rho \right). \quad (25)$$

Integrating the second term by parts once, noting that $\sigma_{ns} = \hbar$ is spatially uniform (i.e., $\partial_q \hbar = 0$), and that $\rho = R_Q^2$, we obtain:

$$I = -\frac{\hbar^2}{2} \int dq \left( g^{ij} \frac{\partial_{q_i} \partial_{q_j} R_Q}{R_Q} \rho + \frac{\partial_{q_i} g^{ij}}{2} \partial_{q_j} \rho \right)$$
$$= -\frac{\hbar^2}{2} \int dq \left( g^{ij} \frac{\partial_{q_i} \partial_{q_j} R_Q}{R_Q} \rho + \partial_{q_i} g^{ij} \frac{\partial_{q_j} R_Q}{R_Q} \rho \right). \quad (26)$$

Substituting back into Eq. (23) yields

$$\langle \mathcal{O} \rangle_{\{S_Q, \rho\}} = \int dq \left[ \frac{g^{ij}}{2} \left( \partial_{q_i} S_Q - A_i \right) \left( \partial_{q_j} S_Q - A_j \right) + V - \frac{\hbar^2}{2} \left( g^{ij} \frac{\partial_{q_i} \partial_{q_j} R_Q}{R_Q} + \partial_{q_i} g^{ij} \frac{\partial_{q_j} R_Q}{R_Q} \right) \right] \rho. \quad (27)$$

We can show that Eq. (27), which is the ensemble average of $\mathcal{O}(p, q)$ of Eq. (20) within the ontological model, is exactly equal to the quantum mechanical expectation value given by the right hand side of Eq. (18), as mentioned in Theorem 2. To do this, let us calculate the quantum mechanical expectation value of the following Hermitian operator (quantum observable):

$$\hat{\mathcal{O}} = (1/2)(\hat{p}_i - A_i(\hat{q})) g^{ij}(\hat{q}) (\hat{p}_j - A_j(\hat{q})) + V(\hat{q}), \quad (28)$$

over a quantum state $|\psi\rangle$. Note that we have chosen a specific ordering in which $g^{ij}(\hat{q})$ is sandwiched between two operators $(\hat{p}_i - A_i(\hat{q}))$. In the position representation, writing $\langle q_i | \hat{p}_i | q_i' \rangle = -i\hbar \partial_{q_i} \delta(q_i - q_i')$, we have to compute

$$\langle \psi | \hat{\mathcal{O}} | \psi \rangle = \int dq\, \psi^* \left[ \frac{1}{2} (-i\hbar \partial_{q_i} - A_i) g^{ij}(q) (-i\hbar \partial_{q_j} - A_j) + V \right] \psi. \quad (29)$$

Now inserting the wave function in Eq. (19), namely $\psi = R_Q \exp(iS_Q/\hbar)$, $R_Q = \sqrt{\rho}$, evaluating the spatial differentiations straightforwardly, and again recalling that $\sigma_{ns} = \hbar$ is spatially uniform, we can divide the integral into real and imaginary parts $I_r$ and $I_i$:

$$\langle \psi | \hat{\mathcal{O}} | \psi \rangle = I_r + I_i, \quad (30)$$

where the real part is

$$I_r = \int dq \left[ \frac{g^{ij}}{2} \left( \partial_{q_i} S_Q - A_i \right) \left( \partial_{q_j} S_Q - A_j \right) + V - \frac{\hbar^2}{2} \left( g^{ij} \frac{\partial_{q_i} \partial_{q_j} R_Q}{R_Q} + \partial_{q_i} g^{ij} \frac{\partial_{q_j} R_Q}{R_Q} \right) \right] \rho, \quad (31)$$

and, since $g^{ij} = g^{ji}$, the imaginary part is

$$I_i = i\frac{\hbar}{2} \int dq \left[ -g^{ij} \partial_{q_i} R_Q^2 \partial_{q_j} S_Q - \partial_{q_i} g^{ij} R_Q^2 \partial_{q_j} S_Q - g^{ij} R_Q^2 \partial_{q_i} \partial_{q_j} S_Q + g^{ij} \partial_{q_i} R_Q^2 A_j + \partial_{q_i} g^{ij} R_Q^2 A_j + g^{ij} R_Q^2 \partial_{q_i} A_j \right]. \quad (32)$$

Indeed, the real part $I_r$ already equals Eq. (27). We only need to check that the imaginary part $I_i$ vanishes. Note that the integral by parts of the third term on the right hand side of Eq. (32) cancels the first and second terms. Also, integrating the sixth term by parts cancels the fourth and fifth terms. Hence $I_i$ vanishes. Of course it does, since $\hat{\mathcal{O}}$ as defined in Eq. (28) is a Hermitian operator, hence the quantum mechanical expectation value must be real.

**An alternative derivation of the uncertainty relation.** In the main text, the uncertainty relation of Eq. (10) is derived via Theorem 2. (See Eq. 8.) Here we show that the uncertainty relation is in fact directly implied by the choice of the epistemic restriction of Eqs. (6) and (7).

Let us consider a pair of canonical conjugate variables corresponding to the $i$-th degree of freedom $(p_i, q_i)$. First, the normalization condition $\int dq \rho(q) = 1$ of the probability density $\rho(q)$ can be written, via integration by parts, as

$$-1 = \int dq\, (q_i - q_{0_i}) \partial_{q_i} \rho = \int dq (q_i - q_{0_i}) \sqrt{\rho} \frac{\partial_{q_i} \rho}{\sqrt{\rho}},$$

where $q_{0_i}$ is an arbitrary number. Applying the Cauchy–Schwartz inequality to the integral on the right, we find

$$\int dq\, (q_i - q_{0_i})^2 \rho(q) \int dq \left( \frac{\partial_{q_i} \rho}{\rho} \right)^2 \rho(q) \geq 1. \quad (33)$$

Now we choose $q_{0_i} = \int dq\, q_i \rho(q) \equiv \langle q_i \rangle_{\{S_Q, \rho\}}$ and we write $\sigma_{q_i}^2 \equiv \int dq (q_i - \langle q_i \rangle_{\{S_Q, \rho\}})^2 \rho(q)$. Multiplying both sides of Eq. (33) by $\sigma_{ns}^2/4$ and recalling that $\sigma_{ns}$ is independent of $q$, we get:

$$\sigma_{q_i}^2 \int dq \left( \frac{\sigma_{ns}}{2} \frac{\partial_{q_i} \rho}{\rho} \right)^2 \rho(q) \geq \frac{\sigma_{ns}^2}{4}. \quad (34)$$

On the other hand, the variance of $p_i$ at any time can be evaluated as:

$$\sigma_{p_i}^2 \equiv \int dq\, d\xi\, dp \left( p_i - \langle p_i \rangle_{\{S_Q, \rho\}} \right)^2 P(p|q, \xi, S_Q, \rho) \mu(\xi) \rho(q)$$
$$= \int dq\, d\xi \left( \frac{\xi}{2} \frac{\partial_{q_i} \rho}{\rho} + \left( \partial_{q_i} S_Q - \langle p_i \rangle_{\{S_Q, \rho\}} \right) \right)^2 \mu(\xi) \rho(q)$$
$$= \int dq \left( \frac{\overline{\xi^2}}{4} \right) \left( \frac{\partial_{q_i} \rho}{\rho} \right)^2 \rho(q) + \int dq \left( \partial_{q_i} S_Q - \langle p_i \rangle_{\{S_Q, \rho\}} \right)^2 \rho(q)$$
$$\geq \int dq \left( \frac{\overline{\xi^2}}{4} \right) \left( \frac{\partial_{q_i} \rho}{\rho} \right)^2 \rho(q) = \int dq \left( \frac{\sigma_{ns}}{2} \frac{\partial_{q_i} \rho}{\rho} \right)^2 \rho(q), \quad (35)$$

where from the first to the second line we have used Eq. (6), and to get the third line we have imposed Eq. (7). Finally, multiplying both sides of Eq. (35) by $\sigma_{q_i}^2$ and using Eq. (34), one obtains

$$\sigma_{q_i} \sigma_{p_i} \geq \frac{\sigma_{ns}}{2} = \frac{\hbar}{2}, \quad (36)$$

where we have again used Eq. (7) that $\sigma_{ns} = \hbar$.

**Proof of Theorem 3. Theorem 3:** Consider an ensemble of trajectories satisfying Eqs. (6) and (7), where $\sigma_{ns} = \hbar$ is constant in space and time. Given a classical Hamiltonian $H(p, q)$ that is at most quadratic in p, and an ensemble of trajectories conserving the average energy and probability current, there is a unique time evolution for $\psi$ given by the (linear and unitary) Schrödinger equation:

$$i\hbar \frac{d}{dt}|\psi\rangle = \hat{H}|\psi\rangle, \qquad (37)$$

where $\hat{H}$ is a Hermitian operator having the same form as that obtained by applying the Dirac canonical quantization procedures to $H(p,q)$ with a specific ordering of operators.

*Proof.* Again, we prove the theorem by considering an ensemble for a single particle of mass $m$ moving in three dimensions in time-independent scalar and vector potentials $V(q)$ and $A(q) = (A_1, A_2, A_2)$, so that the classical Hamiltonian is given by Eq. (12). First, from Eq. (12), the velocity field is related to the momentum as $\dot{q}_i(p) = \partial H/\partial p_i = (p_i - A_i)/m$, so that noting Eqs. (6) and (7), we obtain the average velocity field over the fluctuations of $\xi$ as $\overline{\dot{q}}_i(q;t) \equiv \int d\xi d p \dot{q}_i (p_i) P(p|q,\xi,S_Q,\rho) \mu(\xi) = \int d\xi d p_i (p_i - A_i) \delta(p_i - [\partial_{q_i}S_Q + (\xi/2)\partial_{q_i}\rho/\rho]) \mu(\xi)/m = (\partial_{q_i}S_Q - A_i)/m$. In this case, the assumption of conservation of probability current (no creation or annihilation of trajectories) implies that $\rho(q;t)$ satisfies the following continuity equation:

$$\partial_t \rho + \sum_i \frac{1}{m} \partial_{q_i} \left( \rho (\partial_{q_i}S_Q - A_i) \right) = 0. \qquad (38)$$

On the other hand, the conservation of average energy requires the ensemble of trajectories to satisfy the following equation:

$$\frac{d}{dt} \langle H \rangle_{\{S_Q, \rho\}} = 0, \qquad (39)$$

where $\langle H \rangle_{\{S_Q, \rho\}}$ is the ensemble (phase space) average of the classical energy within the ontological model defined in Eq. (8).

We impose the constraints of conservation of probability current and average energy (Eqs. (38) and (39)) to the dynamics of the ensemble of trajectories. First, from Eqs. (6) and (12), the ensemble average of energy is

$$\begin{aligned}\langle H \rangle_{\{S_Q, \rho\}} &= \int dq d\xi d p \left[ \sum_i \frac{(p_i - A_i(q))^2}{2m} + V(q) \right] P(p|q,\xi,S_Q,\rho)\mu(\xi)\rho(q)\\ &= \int dq d\xi \left[ \sum_i \frac{(\partial_{q_i}S_Q + \xi(\partial_{q_i}\rho/2\rho - A_i)^2}{2m} + V \right] \mu(\xi)\rho(q)\\ &= \int dq d\rho \left[ \sum_i \frac{(\partial_{q_i}S_Q - A_i)^2}{2m} + \frac{\hbar^2}{8m}\left(\frac{\partial_{q_i}\rho}{\rho}\right)^2 + V \right], \end{aligned} \qquad (40)$$

where the second line follows from a trivial integration over $p$ and the third line is due to Eq. (7). Differentiating with respect to time and assuming that $\sigma_{ns} = \hbar$ is constant in time, we get

$$\begin{aligned}\frac{d}{dt}\langle H \rangle_{\{S_Q, \rho\}} &= \int dq \, (\partial_t \rho)\left[ \sum_i \frac{(\partial_{q_i}S_Q - A_i)^2}{2m} + \frac{\hbar^2}{8m}\left(\frac{\partial_{q_i}\rho}{\rho}\right)^2 + V \right]\\ &+ \int dq \rho \left[ \sum_i \frac{(\partial_{q_i}S_Q - A_i)}{m}\partial_{q_i}\partial_t S_Q + \frac{\hbar^2}{4m}\left(\frac{\partial_{q_i}\rho}{\rho}\right)\partial_{q_i}\left(\frac{\partial_t \rho}{\rho}\right) \right].\end{aligned} \qquad (41)$$

Integrating by parts the two terms in the second line, noting that $\partial_q \hbar = 0$, and using Eq. (38), we can rewrite Eq. (41) as:

$$\frac{d}{dt}\langle H \rangle_{\{S_Q, \rho\}} = \int dq \, \partial_t \rho \left( \partial_t S_Q + \sum_i \frac{(\partial_{q_i}S_Q - A_i)^2}{2m} + \frac{\hbar^2}{2m}\left[ \frac{1}{4}\left(\frac{\partial_{q_i}\rho}{\rho}\right)^2 - \frac{1}{2}\frac{\partial_{q_i}^2\rho}{\rho} \right] + V \right). \qquad (42)$$

Using the identity of Eq. (24) and imposing the requirement of the conservation of average energy (Eq. 39) we obtain:

$$\int dq \, \partial_t \rho \left( \partial_t S_Q + \sum_i \frac{(\partial_{q_i}S_Q - A_i)^2}{2m} - \frac{\hbar^2}{2m}\frac{\partial_{q_i}^2 R_Q}{R_Q} + V \right) = 0. \qquad (43)$$

To be valid for any $\partial_t \rho$, the term inside the bracket in the integrand of the above equation must vanish identically. We finally get

$$\partial_t S_Q + \sum_i \frac{(\partial_{q_i}S_Q - A_i)^2}{2m} - \frac{\hbar^2}{2m}\frac{\partial_{q_i}^2 R_Q}{R_Q} + V = 0. \qquad (44)$$

We have thus a pair of coupled Eqs. (38) and (44) which govern the time evolution of $\rho(q;t)$ and $S_Q(q;t)$, respectively, arising from the assumption of conservation of probability current and conservation of average energy. Using the definition of the wave function given by Eq. (9), noting that $R_Q = \sqrt{\rho}$ and bearing in mind the assumption that $\sigma_{ns} = \hbar$ is constant in space and time, we can recast Eqs. (38) and (44) into the following compact form:

$$i\hbar\partial_t \psi = \left( \sum_i \frac{1}{2m}\left(-i\hbar\partial_{q_i} - A_i\right)^2 + V \right)\psi. \qquad (45)$$

Equation (45) is just the familiar Schrödinger equation in the position representation, for a quantum particle of mass $m$ in three-dimensional space subject to a scalar potential $V(q)$ and a vector potential $A(q) = (A_1, A_2, A_3)$ with a Hermitian quantum Hamiltonian $\hat{H} = \sum_i (\hat{p}_i - A_i(\hat{q}))^2/(2m) + V(\hat{q})$.

This derivation of the Schrödinger equation closely parallels the derivation of the classical Hamilton–Jacobi equation given in "Methods" subsection "Proof of Theorem 1." Both fundamental equations are developed within the same framework and are singled out by imposing two common axioms—conservation of average energy and conservation of trajectories (probability current). It is easy to check that the only difference is to derive the Hamilton–Jacobi equation one starts with Eq. (3), whereas to derive the Schrödinger equation one has to replace the classical kinematics of Eq. (3) with Eq. (6). As discussed in "Results" subsection "Microscopic ontic extension and epistemic restriction" and in "Methods" subsection "Statistical interpretation of the epistemic restriction", Eq. (6) can be interpreted as the manifestation of an epistemic restriction which is absent in the classical mechanics.

Exactly the same framework and derivations apply to many-particle systems with a classical general Hamiltonian up to second order in momentum, as spelled out in Theorem 3. In particular, when there are interactions among degrees of freedom, the fundamental nonseparability of $\xi$ will play a crucial role. An example of special interest—because it generates entanglement—appears in the next subsection. Various methods for deriving Schrödinger's equation are also reported in refs. 42,44–63. Our method is distinguished in its modification of classical statistical mechanics via an ontic extension (introducing a global-nonseparable random variable $\xi$), and a specific form of an epistemic restriction (Eq. 6), and imposing the principles of conservation of average energy and conservation of probability current. We emphasize that without the ontic extension and the epistemic restriction, our derivation yields the classical Hamilton–Jacobi equation.

**Schrödinger's equation for measurement of angular momentum.** Let us apply our model to derive the Schrödinger equation for a measurement of angular momentum, adopting von Neumann's prescription for a measurement interaction. We will follow all the steps of the previous subsection. For simplicity, let us model the measurement setup via two interacting particles in positions denoted by $q_S$ and $q_\Sigma$; they represent, respectively, the position of the measured system and the position of the pointer of a measuring device. Let us denote the corresponding conjugate momentum $p_S$ and $p_\Sigma$. Without loss of generality, we consider a measurement of the $z$-component of angular momentum, $l_{z_S} = (q_S \times p_S)_z = x_S p_{y_S} - y_S p_{x_S}$, where $q_S \equiv (x_S, y_S, z_S)$ and $p_S \equiv (p_{x_S}, p_{y_S}, p_{z_S})$. The classical Hamiltonian corresponding to the von Neumann interaction is then

$$H_I = g l_{z_S} p_\Sigma = g (x_S p_{y_S} - y_S p_{x_S}) p_\Sigma, \qquad (46)$$

where $g$ is the interaction coupling. For simplicity, we take $g$ to be constant during the measurement; otherwise $g = 0$.

Taking this interaction Hamiltonian $H_I$ to express the velocity in term of momentum via $\dot{q}_i(p) = \partial H/\partial p_i$ and using Eqs. (6) and (7), we obtain the velocity field averaged over the fluctuations of $\xi$, i.e., $\overline{\dot{q}}_i(q;t) \equiv \int d\xi dp \dot{q}_i(p) P(p|q,\xi,S_Q,\rho)\mu(\xi)$, which is given by:

$$\overline{\dot{x}}_S = -g y_S \partial_{q_\Sigma}S_Q, \quad \overline{\dot{y}}_S = g x_S \partial_{q_\Sigma}S_Q, \quad \overline{\dot{q}}_\Sigma = g \left( x_S \partial_{y_S}S_Q - y_S \partial_{x_S}S_Q \right), \qquad (47)$$

and $\overline{\dot{z}}_S = 0$. Assuming that the probability current is conserved, we arrive at the following continuity equation:

$$\begin{aligned}\partial_t \rho &- g y_S \partial_{x_S}\left( \rho \partial_{q_\Sigma}S_Q \right) + g x_S \partial_{y_S}\left( \rho \partial_{q_\Sigma}S_Q \right) + g x_S \partial_{q_\Sigma}\left( \rho \partial_{y_S}S_Q \right)\\ &- g y_S \partial_{q_\Sigma}\left( \rho \partial_{x_S}S_Q \right) = 0.\end{aligned} \qquad (48)$$

Now let us impose conservation of average energy. First, using Eq. (46) the ensemble of energy reads

$$\begin{aligned}\langle H_I \rangle_{\{S_Q, \rho\}} = &\int dq d\xi dp g (x_S p_{y_S} - y_S p_{x_S}) p_\Sigma\\ &\times P(p_{x_S}, p_{y_S}, p_\Sigma | q, \xi, S_Q, \rho)\mu(\xi)\rho(q).\end{aligned} \qquad (49)$$

Inserting Eq. (6) and evaluating the integrations over $p$ and $\xi$, we get, noting Eq. (7),

$$\begin{aligned}\langle H_I \rangle_{\{S_Q, \rho\}} = \int dq \Big\{ &g (x_S \partial_{y_S}S_Q - y_S \partial_{x_S}S_Q) \partial_{q_\Sigma}S_Q \rho\\ &+ g \frac{\hbar^2}{4}\left[ x_S \left(\frac{\partial_{y_S}\rho}{\rho}\right)\left(\frac{\partial_{q_\Sigma}\rho}{\rho}\right) - y_S\left(\frac{\partial_{x_S}\rho}{\rho}\right)\left(\frac{\partial_{q_\Sigma}\rho}{\rho}\right) \right] \rho \Big\}.\end{aligned} \qquad (50)$$

Taking the derivative with respect to time to both sides, noting that $\sigma_{ns} = \hbar$ is constant in time, we obtain, after a long but straightforward calculation, and rearrangement,

$$\begin{aligned}\frac{d}{dt}\langle H_I \rangle_{\{S_Q, \rho\}} = \int dq \Big\{ &(\partial_t \rho) g (x_S \partial_{y_S}S_Q - y_S \partial_{x_S}S_Q)\partial_{q_\Sigma}S_Q\\ &+ (g y_S \partial_{x_S}(\rho \partial_{q_\Sigma}S_Q) - g x_S \partial_{y_S}(\rho \partial_{q_\Sigma}S_Q)\\ &- g x_S \partial_{q_\Sigma}(\rho \partial_{y_S}S_Q) + g y_S \partial_{q_\Sigma}(\rho \partial_{x_S}S_Q)) \partial_t S_Q\\ &+ g \hbar^2 \Big[ x_S \left( \frac{1}{4}\left(\frac{\partial_{y_S}\rho}{\rho}\right)\left(\frac{\partial_{q_\Sigma}\rho}{\rho}\right) - \frac{\partial_{y_S}\partial_{q_\Sigma}\rho}{2\rho} \right)\\ &- y_S \left( \frac{1}{4}\left(\frac{\partial_{x_S}\rho}{\rho}\right)\left(\frac{\partial_{q_\Sigma}\rho}{\rho}\right) - \frac{\partial_{x_S}\partial_{q_\Sigma}\rho}{2\rho} \right) \Big] \partial_t \rho \Big\}.\end{aligned} \qquad (51)$$

 

Here, bearing in mind that $\hbar$ is constant of space, we have performed partial integrations where appropriate. Using Eq. (48), the second line can be simplified into $\partial_q \rho \partial_s S_C$; moreover the last line can be simplified by virtue of the identity Eq. (24), so that the whole equation simplifies into

$$\frac{d}{dt} \langle H_1 \rangle_{\{S_Q,\varphi\}} = \int dq \, (\partial_t \rho) \left\{ \partial_t S_Q + g \left( x_S \partial_{y_S} S_Q - y_S \partial_{x_S} S_Q \right) \partial_{q_\Sigma} S_Q \right. \\ \left. - g\hbar^2 \left( x_S \frac{\partial_{x_S} \partial_{q_\Sigma} R_Q}{R_Q} - y_S \frac{\partial_{y_S} \partial_{q_\Sigma} R_Q}{R_Q} \right) \right\}, \quad (52)$$

where $R_Q \equiv \sqrt{\rho}$. Conservation of average energy, $(d/dt) \langle H_1 \rangle_{\{S_Q,\varphi\}} = 0$ for any $\partial_t \rho$, means that the integrand inside the bracket must vanish. So we have

$$\partial_t S_Q + g \left( x_S \partial_{y_S} S_Q - y_S \partial_{x_S} S_Q \right) \partial_{q_\Sigma} S_Q - g\hbar^2 \left( x_S \frac{\partial_{y_S} \partial_{q_\Sigma} R_Q}{R_Q} - y_S \frac{\partial_{x_S} \partial_{q_\Sigma} R_Q}{R_Q} \right) = 0. \quad (53)$$

We thus have a pair of coupled Eqs. (48) and (53), arising, respectively, from the conservation of probability current and conservation of average energy. Finally, applying the definition $\psi = R_Q \exp(iS_Q/\hbar)$ of wave function in Eq. (9), and noting that $\sigma_{ns} = \hbar$ is constant in space and time, we recast Eqs. (48) and (53) into

$$i\hbar \frac{d}{dt} |\psi\rangle = \hat{H}_1 |\psi\rangle. \quad (54)$$

Here $\hat{H}_1$ is a Hermitian operator defined as:

$$\hat{H}_1 \equiv g \hat{l}_{z_S} \hat{p}_\Sigma, \quad (55)$$

where $\hat{p}_i$ is the quantum momentum operator for the $i$-degree of freedom and $\hat{l}_{z_S} \equiv \hat{x}_S \hat{p}_{y_S} - \hat{y}_S \hat{p}_{x_S}$ is the $z$-component of the quantum angular momentum operator of the measured system. This equation is just the Schrödinger equation for a measurement of angular momentum via the von Neumann measurement interaction $\hat{H}_1$.

In this derivation of Schrödinger's equation for a measurement interaction, the nonseparability of $\xi$ plays a crucial role. To see this, let us instead suppose that $\xi$ is separable into three random variables $\xi = (\xi_{x_S}, \xi_{y_S}, \xi_{q_\Sigma})$ each associated with the respective degrees of freedom $(x_S, y_S, q_\Sigma)$, with vanishing average $\overline{\xi}_{x_S} = \overline{\xi}_{y_S} = \overline{\xi}_{q_\Sigma} = 0$, so that the pairs $(\xi_{x_S}, \xi_{q_\Sigma})$ and $(\xi_{y_S}, \xi_{q_\Sigma})$ were both independent of each other (thus uncorrelated): $\overline{\xi_{x_S} \xi_{q_\Sigma}} = \overline{\xi}_{x_S} \overline{\xi}_{q_\Sigma} = 0 = \overline{\xi}_{y_S} \overline{\xi}_{q_\Sigma} = \overline{\xi_{y_S} \xi_{q_\Sigma}}$. In this case, the last term in Eq. (50) (explicitly proportional to $\hbar^2$) vanishes, yielding

$$\langle H_1 \rangle_{\{S_Q,\varphi\}} = \int dq \, g \left( x_S \partial_{y_S} S_Q - y_S \partial_{x_S} S_Q \right) \partial_{q_\Sigma} S_Q \rho. \quad (56)$$

Identifying $S_Q$ as the classical Hamilton's principal function $S_C$, and recalling Eq. (3), the above expression is just the conventional classical average energy.

Then, imposing the conservation of average energy $(d/dt) \langle H_1 \rangle_{\{S_Q,\varphi\}} = 0$ with $\langle H_1 \rangle_{\{S_Q,\varphi\}}$ given by Eq. (56) and using Eq. (48), instead of Eq. (53) we obtain

$$\partial_t S_Q + g \left( x_S \partial_{y_S} S_Q - y_S \partial_{x_S} S_Q \right) \partial_{q_\Sigma} S_Q = 0. \quad (57)$$

This is just the classical Hamilton–Jacobi equation which can be seen by identifying $S_Q$ as the Hamilton's principal function $S_C$, and noting Eqs. (1) and (46). Notice that, comparing Eq. (57) with Eq. (53), the last term in Eq. (53) which explicitly depends on $\hbar^2$ (obtained with $\xi$ nonseparable), is no longer present in Eq. (57) (obtained with separable $\xi$). This $\hbar^2$-dependent term is called quantum potential in Bohmian mechanics, and is generally argued as being responsible for the classically puzzling quantum phenomena. Hence, the fundamental nonseparability of $\xi$ plays a crucial role in the derivation of the Schrödinger equation for interacting systems. Since such interaction implies quantum entanglement—for the above example for the measurement interaction, we obtain, in the next subsection, entanglement between the system and the apparatus—the nonseparability of $\xi$ is indeed crucial for obtaining quantum entanglement.

**Derivation of Born's rule**. Let us apply our ontological model to a measurement of $\mathcal{O}_S(p_S, q_S)$ using a von Neumann measurement interaction Hamiltonian $H_I = g\mathcal{O}_S p_\Sigma$. Here $p_\Sigma$ is the momentum of the pointer on a measuring device, conjugate to the pointer position $q_\Sigma$, and $g$ is the interaction coupling. As spelled out in Theorem 3, the resulting Schrödinger equation reads $i\hbar(d/dt) |\psi\rangle = \hat{H}_1 |\psi\rangle$, where the quantum Hamiltonian is $\hat{H}_1 = g\hat{\mathcal{O}}_S \hat{p}_\Sigma$. An example of the derivation of the Schrödinger equation for the measurement of angular momentum is given in the previous subsection. From the Schrödinger equation governing the time evolution of the wave function during the measurement interaction, we can proceed to describe measurements reproducing the predictions of quantum mechanics as prescribed by Born's rule, as follows.

We let $\psi_S(q_S)$ denote the wave function of the system at the initial measurement interaction time $t = 0$. It can be expanded as $\psi_S(q_S) = \sum_k c_k \phi_{S_k}(q_S)$, where $\{|\phi_{S_k}\rangle\}$, $k = 0, 1, 2, \ldots$ is the complete set of orthonormal eigenvectors of the Hermitian operator $\hat{\mathcal{O}}_S$ with the corresponding eigenvalues $\{o_k\}$, satisfying $\hat{\mathcal{O}}_S |\phi_{S_k}\rangle = o_k |\phi_{S_k}\rangle$. The expansion coefficient is then $c_k = \int dq_S \phi^*_{S_k}(q_S) \psi_S(q_S) = \langle \phi_{S_k} | \psi_S \rangle$. Let $\varphi_\Sigma(q_\Sigma)$ denote the initial wave function of the pointer of a measuring

device, and assume that the total wave function of the system and device at $t = 0$ is factorizable: $\psi(q_S, q_\Sigma; 0) = \psi_S(q_S) \varphi_\Sigma(q_\Sigma) = \sum_k c_k \phi_{S_k}(q_S) \varphi_\Sigma(q_\Sigma)$. It evolves in time in accordance with Schrödinger's equation $i\hbar(d/dt) |\psi\rangle = \hat{H}_1 |\psi\rangle$ with the measurement-interaction quantum Hamiltonian $\hat{H}_1 = g\hat{\mathcal{O}}_S \hat{p}_\Sigma$; thus the total wave function at the end of measurement interaction at time $t = T$ is entangled:

$$\psi(q_S, q_\Sigma; T) = \sum_k c_k \phi_{S_k}(q_S) \varphi_\Sigma(q_\Sigma - g o_k T). \quad (58)$$

Let us assume that the strength of the interaction coupling $g$ is sufficient such that at $t = T$ the series of device wave packets $\{\varphi_\Sigma(q_\Sigma - g o_j T)\}$, $j = 0, 1, 2, \ldots$, for different values of $j$, effectively do not overlap. If so, when the position of the pointer of the device $q_\Sigma$ at the end of the measurement process belongs to the support of $\varphi_\Sigma(q_\Sigma - g o_j T)$, we can unambiguously register the outcome of measurement as $o_j$, one of the eigenvalues of $\hat{\mathcal{O}}_S$. The probability that the measurement yields $o_j$ given $q = (q_S, q_\Sigma)$ and $\psi(T)$ is thus

$$P(o_j | q_S, q_\Sigma, \psi(T)) = 1 \{q_\Sigma \in \Lambda_j\}, \quad (59)$$

where $\Lambda_j$ is the support of $\varphi_\Sigma(q_\Sigma - g o_j T)$ and $1\{"event"\}$ is an indicator function which gives "1" if the event occurs and "0" if not.

Furthermore, from the definition of the wave function of Eq. (9) and Eq. (58), the probability that the configuration of the system and device at $t = T$ is $q = (q_S, q_\Sigma)$ is given by:

$$P(q_S, q_\Sigma | \psi(T)) = |\psi(q_S, q_\Sigma; T)|^2 \\ = \sum_{(j,k)} c^*_j c_k \phi^*_{S_j}(q_S) \phi_{S_k}(q_S) \varphi^*_\Sigma(q_\Sigma - g o_j T) \varphi_\Sigma(q_\Sigma - g o_k T) \\ = \sum_k |c_k|^2 |\phi_{S_k}(q_S)|^2 |\varphi_\Sigma(q_\Sigma - g o_k T)|^2, \quad (60)$$

where in the last equality we have taken into account the fact that since the support of $\{\varphi_\Sigma(q_\Sigma - g o_k T)\}$ for different $k$ do not overlap, the cross terms in the double sum all vanish.

One can finally show straightforwardly from Eqs. (59) and (60), via conventional probability theory, that the probability to get $o_j$ when the initial wave function of the system is $\psi_S = \sum_k c_k \phi_{S_k}$ is given by the celebrated Born's rule:

$$P(o_j | \psi_S) = \int dq_S dq_\Sigma P(o_j | q_S, q_\Sigma, \psi(T)) P(q_S, q_\Sigma | \psi(T)) \\ = \int dq_S dq_\Sigma 1 \{q_\Sigma \in \Lambda_j\} \sum_k |c_k|^2 |\phi_{S_k}(q_S)|^2 |\varphi_\Sigma(q_\Sigma - g o_k T)|^2 \\ = \int dq_S dq_\Sigma |c_j|^2 |\phi_{S_j}(q_S)|^2 |\varphi_\Sigma(q_\Sigma - g o_j T)|^2 = |c_j|^2 = |\langle \phi_{S_j} | \psi_S \rangle|^2, \quad (61)$$

where in the last line we have used the normalizations of $\phi_{S_k}(q_S)$ and $\varphi_\Sigma(q_\Sigma - g o_j T)$ as implied by the definitions given in Eq. (9).

Suppose that the position of the pointer on the measuring device belongs to the support of $\varphi_\Sigma(q_\Sigma - g o_j T)$, i.e., that the outcome of measurement is $o_j$. If the measurement is not destructive and since $\varphi_\Sigma(q_\Sigma - g o_j T)$ for different $j$ do not overlap, Eq. (58) implies that the "effective" wave function of the system and device becomes $\phi_{S_j}(q_S) \varphi_\Sigma(q_\Sigma - g o_j T)$. Hence, when the outcome of measurement is $o_j$, the effective wave function of the system alone is $\phi_{S_j}(q_S)$, i.e., the eigenstate of $\hat{\mathcal{O}}_S$ associated with the eigenvalue $o_j$.

In this connection, Wallstrom[29] has argued that a derivation of the Schrödinger equation based on the combination of a modified Hamilton–Jacobi equation and a continuity equation via Eq. (9), as in our model (in the previous two subsections), will have to allow many more wave functions than those allowed in quantum mechanics. In particular, the wave function $\psi$ defined in Eq. (9) is in general not single-valued (since the phase function $S_Q$ is in general many-valued, for example for wave functions with angular momentum). This is also the feature of Nelson's stochastic mechanics[46] and many other approaches[42,44,45,47]. He went on to argue that unless one imposes, by hand, a quantization condition as in the old quantum theory (to ensure the single-valuedness of the wave function), then one has, for example, to allow a particle to have a non-integral (continuum) value of angular momentum. Such an ad hoc condition will, in our model, physically translate into an additional statistical constraint which selects a yet narrower class of ensembles of trajectories.

Within our model, by construction, a particle may indeed have a non-integral value of angular momentum (or a continuum value of energy) if left unmeasured. In this case, the wave function may indeed not be single-valued. Nevertheless, as shown above, measurement of angular momentum will only yield discrete (quantized) outcome as in quantum mechanics. Hence, discrete quantum numbers is an emergent feature of measurement, rather than an objective property of the system regardless of measurement. Remarkably, as shown by Theorem 2 and above in this subsection, the ensemble average of the angular momentum prior to measurement (in which the angular momentum may take continuum values) is well defined and is equal to the quantum mechanical expectation value obtained in measurement (in which each single shot yields discrete integral value). A similar answer to Wallstrom's objection is argued in refs. [45,47].

## References

1. Einstein, A., Podolsky, B. & Rosen, N. Can quantum-mechanical description of physical reality be considered complete? *Phys. Rev.* **47**, 777–780 (1935).
2. Howard, D. in *Sixty-Two Years of Uncertainty: Historical, Philosophical and Physical Inquiries into the Foundations of Quantum Mechanics* (ed. Miller, A.) 61–111 (Plenum, New York, 1990).
3. Ballentine, L. E. The statistical interpretation of quantum mechanics. *Rev. Mod. Phys.* **42**, 358–381 (1970).
4. Peierls, R. *Surprises in Theoretical Physics* (Princeton University Press, Princeton, 1979).
5. Jaynes, E. T. *Foundations of Radiation Theory and Quantum Electrodynamics* (Plenum, New York, 1980).
6. Emerson, J. V. *Quantum Chaos and Quantum-Classical Correspondence* (PhD dissertation, Simon Fraser Univ., 2001).
7. Spekkens, R. W. Evidence for the epistemic view of quantum states: a toy theory. *Phys. Rev. A* **75**, 032110 (2007).
8. Harrigan, N. & Spekkens, R. W. Einstein, incompleteness, and the epistemic view of quantum states. *Found. Phys.* **40**, 125–157 (2010).
9. Hardy, L. Disentangling nonlocality and teleportation. Preprint at https://arxiv.org/abs/quant-ph/9906123 (1999).
10. Van Enk, S. J. A toy model for quantum mechanics. *Found. Phys.* **37**, 1447–1460 (2007).
11. Bartlett, S. D., Rudolph, T. & Spekkens, R. W. Reconstruction of Gaussian quantum mechanics from Liouville mechanics with an epistemic restriction. *Phys. Rev. A* **86**, 012103 (2012).
12. Spekkens, R. W. in *Quantum Theory: Informational Foundations and Foils* (eds Spekkens, R. W. & Chiribella, G. M.) 83–135 (Fundamental Theories of Physics Vol. 181, Springer, Dordrecht, 2015).
13. Hardy, L. Quantum ontological excess baggage. *Stud. Hist. Phil. Mod. Phys.* **35**, 267–276 (2004).
14. Montina, A. Exponential complexity and ontological theories of quantum mechanics. *Phys. Rev. A* **77**, 022104 (2008).
15. Kochen, S. & Specker, E. P. The problem of hidden variables in quantum mechanics. *J. Math. Mech.* **17**, 59–87 (1967).
16. Spekkens, R. W. Contextuality for preparations, transformations, and unsharp measurements. *Phys. Rev. A* **71**, 052108 (2005).
17. Bohm, D. A suggested interpretation of the quantum theory in terms of "hidden" variables. I. *Phys. Rev* **85**, 166–179 (1952).
18. Dürr, D., Goldstein, S. & Zangh, N. in *Experimental Metaphysics—Quantum Mechanical Studies for Abner Shimony, Volume One* (eds Cohen, R. S., Horne, M. & Stachel, J.) 25–38 (Boston Studies in the Philosophy of Science Vol. 193, Kluwer, Dordrecht, 1997).
19. Pusey, M. F., Barrett, J. & Rudolph, T. On the reality of the quantum state. *Nat. Phys.* **8**, 475–478 (2012).
20. Bell, J. S. On the Einstein-Podolsky-Rosen paradox. *Physics* **1**, 195–200 (1964).
21. Bell, J. S. On the problem of hidden variables in quantum mechanics. *Rev. Mod. Phys.* **38**, 447–452 (1966).
22. Aharonov, Y. & Rohrlich, D. *Quantum Paradoxes: Quantum Theory for the Perplexed* (Wiley-VCH, Weinheim, 2005).
23. Rund, H. *The Hamilton-Jacobi Theory in the Calculus of Variations: Its Role in Mathematics and Physics* (Van Nostrand, London, 1966).
24. Howard, D. Einstein on locality and separability. *Stud. Hist. Phil. Sci.* **16**, 171–201 (1985).
25. Emerson, J., Serbin, D., Sutherland, C. & Veitch, V. The whole is greater than the sum of the parts: on the possibility of purely statistical interpretations of quantum theory. Preprint at https://arxiv.org/abs/1312.1345 (2013).
26. Heisenberg, W. Über den anschaulichen Inhalt der quantentheoretischen Kinematik und Mechanik. *Z. Phys.* **43**, 172–198 (1927).
27. Kennard, E. H. Zur Quantenmechanik einfacher Bewegungstypen. *Z. Phys.* **44**, 326–352 (1927).
28. Budiyono, A. Objective uncertainty relation with classical background in a statistical model. *Phys. A* **392**, 43–47 (2013).
29. Wallstrom, T. Inequivalence between the Schrödinger equation and the Madelung hydrodynamic equations. *Phys. Rev. A* **49**, 1613–1617 (1994).
30. Wheeler, J. A. in *Complexity, Entropy, and the Physics of Information* (ed. Zurek, W. H.) 3–28 (Westview Press, New York, 1990).
31. Popescu, S. & Rohrlich, D. Quantum nonlocality as an axiom. *Found. Phys.* **24**, 379–385 (1994).
32. Caves, C. M. & Fuchs, C. A. Quantum information: how much information is in a state vector? in *The dilemma of Einstein, Podolsky and Rosen—*

*60 years later*. (eds Mann, A. & Revzen, M.) 226–257 (Ann. Israel Phys. Soc. **12** 1996).
33. Zeilinger, A. A foundational principle for quantum mechanics. *Found. Phys.* **29**, 631–643 (1999).
34. Hardy, L. Quantum theory from five reasonable axioms. Preprint at https://arxiv.org/abs/quant-ph/0101012v4 (2001).
35. Fuchs, C. A. in *Decoherence and its Implications in Quantum Computation and Information Transfer: Proceedings of the NATO Advanced Research Workshop, Mykonos Greece, June 25–30, 2000* (eds Gonis, A. & Turchi, P. E. A.) 38–82 (IOS Press, Amsterdam, 2001).
36. Barrett, J. Information processing in generalized probabilistic theories. *Phys. Rev. A* **75**, 032304 (2007).
37. Dakić, B. & Brukner, Č. in *Deep Beauty: Understanding the Quantum World through Mathematical Innovation* (ed. Halvorson, H.) 365–392 (Cambridge University Press, Cambridge, 2011).
38. Paterek, T., Dakić, B. & Brukner, Č. Theories of systems with limited information content. *New J. Phys.* **12**, 053037 (2010).
39. Chiribella, G., D'Ariano, G. M. & Perinotti, P. Informational derivation of quantum theory. *Phys. Rev. A* **84**, 012311 (2011).
40. Masanes, L. & Müller, M. P. A derivation of quantum theory from physical requirements. *New J. Phys.* **13**, 063001 (2011).
41. de la Torre, G., Masanes, L., Short, A. J. & Müller, M. P. Deriving quantum theory from its local structure and reversibility. *Phys. Rev. Lett.* **109**, 090403 (2012).
42. Hall, M. J. W. & Reginatto, M. Schrödinger equation from an exact uncertainty principle. *J. Phys. A* **35**, 3289–3303 (2002).
43. Hall, M. J. W., Deckert, D.-A. & Wiseman, H. M. Quantum phenomena modeled by interactions between many classical worlds. *Phys. Rev. X* **4**, 041013 (2014).
44. Markopoulou, F. & Smolin, L. Quantum theory from quantum gravity. *Phys. Rev. D* **70**, 124029 (2004).
45. Smolin, L. Could quantum mechanics be an approximation to another theory? Preprint at https://arxiv.org/abs/quant-ph/0609109 (2006).
46. Nelson, E. Derivation of the Schrödinger equation from Newtonian mechanics. *Phys. Rev.* **150**, 1079–1085 (1966).
47. de la Peña-Auerbach, L., Valdés-Hernández, A., Cetto, A. M. & Franca, H. M. Genesis of quantum nonlocality. *Phys. Lett. A* **375**, 1720–1723 (2011).
48. Feynman, R. P. Space-time approach to non-relativistic quantum mechanics. *Rev. Mod. Phys.* **20**, 367–387 (1948).
49. Fényes, I. Eine wahrscheinlichkeitstheoretische Begründung und Interpretation der Quantenmechanik. *Z. Phys.* **132**, 81–106 (1952).
50. Weizel, W. Ableitung der Quantentheorie aus einem klassischen, kausal determinierten Modell. *Z. Phys.* **134**, 264–285 (1953).
51. Kershaw, D. Theory of hidden variables. *Phys. Rev.* **136**, B1850–B1856 (1964).
52. de la Peña-Auerbach, L. New formulation of stochastic theory and quantum mechanics. *J. Math. Phys.* **10**, 1620–1630 (1969).
53. Santamato, E. Geometric derivation of the Schrödinger equation from classical mechanics in curved Weyl spaces. *Phys. Rev. D* **29**, 216–222 (1984).
54. Frieden, B. R. Fisher information as the basis for the Schrödinger wave equation. *Am. J. Phys.* **57**, 1004–1008 (1989).
55. Garbaczewski, P. & Vigier, J.-P. Quantum dynamics from the Brownian recoil principle. *Phys. Rev. A* **46**, 4634–4638 (1992).
56. Kaniadakis, G. Statistical origin of quantum mechanics. *Phys. A* **307**, 172–184 (2002).
57. Parwani, R. Information measures for inferring quantum mechanics. *J. Phys. A* **38**, 6231–6237 (2005).
58. Santos, E. On a heuristic point of view concerning the motion of matter: from random metric to Schrödinger equation. *Phys. Lett. A* **352**, 49–54 (2006).
59. Scully, M. O. The time dependent Schrödinger equation revisited I: quantum field and classical Hamilton-Jacobi routes to Schrödinger's wave equation. *J. Phys. Conf. Ser.* **99**, 012019 (2008).
60. Caticha, A. Entropic dynamics, time and quantum theory. *J. Phys. A* **44**, 225303 (2011).
61. Field, J. H. Derivation of the Schrödinger equation from the Hamilton-Jacobi equation in Feynman's path integral formulation of quantum mechanics. *Eur. J. Phys.* **32**, 63–87 (2011).
62. Budiyono, A. Quantization from Hamilton-Jacobi theory with a random constraint. *Phys. A* **391**, 4583–4589 (2012).
63. Schleich, W. P., Greenberger, D. M., Kobe, D. H. & Scully, M. O. Schrödinger equation revisited. *Proc. Natl. Acad. Sci. USA* **110**, 5374–5379 (2013).

## Acknowledgements
We thank the John Templeton Foundation (Project ID 43297) and the Israel Science Foundation (Grant 1190/13) for support. The opinions expressed in this publication do not necessarily reflect the views of the John Templeton Foundation. The authors like to thank the anonymous reviewers for their thorough reading of the manuscript and their constructive and insightful comments and suggestions.

## Author contributions

A.B. conceived the ideas, derived the results of this paper, and wrote a first draft. D.R. helped clarify the ideas and rewrite the drafts.

## Additional information

**Competing interests:** The authors declare no competing financial interests.

**Reprints and permission** information is available online at http://npg.nature.com/reprintsandpermissions/

**Publisher's note:** Springer Nature remains neutral with regard to jurisdictional claims in published maps and institutional affiliations.